\documentclass[%
 amsmath,amssymb,
 aps,
]{revtex4-2}

\usepackage{graphicx,subcaption,ragged2e}
\usepackage{dcolumn}
\usepackage{bm}
\usepackage{tikz}
\usepackage{graphicx}
\usetikzlibrary{positioning}

\usepackage{soul} 
\usepackage{float}  
\usepackage{caption}
\usepackage{subcaption}
\usepackage{hyperref}
\usepackage{multirow}
\usepackage{colortbl}
\usepackage{xcolor}

\newcommand{\eq}{\mathrm{eq}}

\begin{document}

\preprint{APS/123-QED}

\title{Artificial Bottleneck Effect in Large Eddy Simulations}
\author{Mostafa Kamal\footnote{abdelmoh@uci.edu}, Perry L. Johnson\footnote{perry.johnson@uci.edu}}
\affiliation{%
 Department of Mechanical and Aerospace Engineering, University of California, Irvine
}%




\date{\today}

\begin{abstract}
In Navier-Stokes turbulence, a bottleneck effect in the energy cascade near the viscous cutoff causes an overshoot in the energy spectrum, or spectral bump, relative to Komogorov's -5/3 power-law scaling. A similar spectral overshoot occurs in large-eddy simulations (LES) when an eddy viscosity model is used. It is not a viscous phenomenon, but rather is caused by error in the residual stress model. This artificial bottleneck effect in LES leads to an over-prediction of kinetic energy even if a reliable dynamic procedure is used to accurately capture the spectral decay at the cutoff length scale.
Recently, Johnson [2022, \textit{J. Fluid Mech.}, \textbf{934}, A30] introduced a physics-inspired generalization of the concept of spatial filtering that provides a dynamic procedure that does not require a test filter calculation.
In this paper, this method of Stokes flow regularization (SFR) is used alongside fundamental considerations related to kinetic energy to generate a range of LES models 
to explore the artificial bottleneck effect in more detail.
The coefficients for each dynamic model are determined locally, without the need of averaging over homogeneous directions.
The theory directly provides stabilizing elements such as local averaging of coefficients.
\textit{A posteriori} tests of the models in isotropic turbulence are reported, 
demonstrating the robustness of the SFR-based dynamic procedure for a range of model forms and providing a framework for fair comparisons between them in terms of their impact on the bottleneck effect. An effective means of mitigating the bottleneck effect is to introduce a nonlinear gradient component in the residual stress closure, forming a dynamic mixed model. One primary reason for the efficacy of this approach is that the nonlinear gradient model is able to accurately capture aspects of the local structure of the residual stresses, leading to a better representation of energy cascade efficiencies.


\end{abstract}

\maketitle

\section{Introduction}
\label{sec:introduction}

Direct numerical simulation (DNS) of turbulent flows at moderate and high Reynolds numbers is often impractical or even impossible for many turbulent flow applications in science and engineering. 
This situation is likely to continue for the foreseeable future.
Large-eddy simulation (LES) offers a more tractable alternative for computing turbulent flows with the use of a coarser grid and larger time step than required for DNS.  However, LES inevitably involves non-negligible modeling error due to the net force of unresolved fluctuations.

The earliest LES model is the eddy viscosity model of Smagorinsky \cite{Smagorinsky1963} and Lilly \cite{Lilly1967}.
Leonard \cite{Leonard1975} introduced the use of a spatial filtering integral operator as the basis for the LES equations, which was later elaborated by Germano \cite{Germano1992}.
While sufficient to dissipate resolved kinetic energy and stabilize coarse grid solutions, it is well-known that eddy viscosity models do not accurately represent the effects of unresolved fluctuations near the grid scale \cite{Borue1998, Ballouz2018Tensor}.
In spite of this, the popularity of eddy viscosity models is entrenched, because the accuracy of grid-scale statistics and flow topologies are often not paramount when simulating simpler flow regimes (e.g., single-phase, unbounded, non-reacting).
This has motivated the development and use of implicit LES (ILES) models, which tailor the numerical discretization to provide the necessary dissipation while forgoing any explicit calculation of a residual stress tensor \cite{boris1992new, boris1973flux, colella1984piecewise, fureby2002large}.
Indeed, dissipative schemes (based on eddy viscosity or numerical dissipation) can seem effective when the main physical phenomenon to be represented is simply the energy cascade \cite{Onsager1949, Jimenez1995, Johnson2021Role}.
The challenges of using LES to simulate complex flow physics (including wall-bounded, multi-phase, and/or reacting flows), continues to motivate work toward more accurate representations of grid-scale turbulence, where there is still much that can be learned even from simpler flows.

For example, the interaction between the kinetic energy cascade and viscous dissipation is not trivial even for Navier-Stokes turbulence (e.g., DNS). The energy spectrum typically exhibits a characteristic spectral bump between the inertial and viscous range of wavenumbers, where the energy spectrum exceeds the inertial range value. This spectral bump, near $0.1 < k \eta  < 0.2$, is observed across different Reynolds numbers. It is  shown in Ref.\ \cite{donzis2010bottleneck} that the location of the bump is independent of $Re_{\lambda}$, and the height of the bump decreases with increasing $Re_{\lambda}$.
This phenomenon is widely recognized and often linked to a bottleneck effect in the energy cascade as it approaches the viscous cutoff scale ($\eta = \nu^{3/4} \epsilon^{-1/4}$). Additionally, it has been noted that the width of the bottleneck seems to remain constant regardless of the Reynolds number, suggesting that the bottleneck's origin is likely due to a viscous effect rather than an effect from the inertial range.

A number of physical explanations for the spectral bump have been put forward. The bottleneck phenomenon is commonly thought to be linked with the diminishing presence of nonlinearities as this range is approached \cite{mininni2008nonlocal}. Ref.\ \cite{frisch2008hyperviscosity} demonstrated that the EDQNM closure produces a spectral bump associated with the bottleneck effect and suggested a possible link to an incomplete thermalization process.
The connection between the bottleneck effect and incomplete thermalization was demonstrated for Navier-Stokes in Ref.\ \cite{Agrawal2020}.
Another explanation provided in Ref.\ \cite{kurien2004cascade} attributes the bottleneck effect to helicity dynamics. These dynamics slow down the cascade of both energy and helicity, which are the two conserved quantities in turbulence. In a more recent study, Ref.\ \cite{Johnson2021Role} proposes an alternative mechanism for energy transfer, attributing it to vortex thinning in the inverse cascade process observed in two dimensions. Understanding this phenomenon remains challenging, even in simple flows such as homogeneous isotropic turbulence (HIT) with a Newtonian viscous stress tensor. However, little research has been conducted on the bottleneck effect within the context of LES, where (eddy) viscosity-based stress tensors are common. A spectral bump or bottleneck effect in LES would presumably hinder the prediction of residual kinetic energy, motivating possible alternatives to algebraic eddy viscosity models. 

The spatial filtering theory of LES can be used to derive alternatives to the eddy viscosity assumption, including the scale-similarity model \cite{Bardina1980} and the nonlinear gradient model \cite{Clark1979}.
In \textit{a priori} tests (i.e., directly testing the accuracy of the model expression by applying a spatial filter to DNS data), such models provide a stark improvement in accuracy over eddy viscosity models \cite{Borue1998, liu1994properties, vreman1995priori}.
However, the scale-similarity or nonlinear gradient models tend to perform poorly in \textit{a posteriori} tests (i.e., applying the model within an LES calculation and evaluating the simulation result) due to insufficient dissipation \cite{vreman1997large}.
Additionally, deconvolution methods have been used to derive closure forms by assuming the filter to be invertible \cite{stolz1999approximate}. While these models exhibit realistic  results in \textit{a priori} testing, they tend to produce nonphysical behavior in \textit{a posteriori} tests due to the fact that they often do not provide enough dissipation.
The result is that structural models tend to require an extra dissipation term, either an explicit model component such as an eddy viscosity or an implicit dissipation via truncation error in the numerical scheme.

Broadly speaking, such mixed models can offer an attractive trade-off, with structural models providing more accuracy in the representation of fluctuations near the grid scale while the dissipative component can bring stability and robustness.
In fact, Johnson \cite{Johnson2021Role} demonstrated that an eddy viscosity approximation more accurately represents multi-scale interactions, and thus a mixed model consisting of a nonlinear gradient model together with an eddy viscosity has a strong basis in turbulence physics.

Another application of spatial filtering theory to improve the accuracy of LES modeling is the Germano identity \cite{Germano1992}, which offers a robust procedure for determining unknown model coefficients. 
A particularly popular instance of this approach uses the Smagorinsky-Lilly eddy viscosity model with dimensionless coefficient determined on the fly by the Germano-Lilly dynamic procedure \cite{Germano1991, Lilly1992}. 
Such dynamic procedures for obtaining coefficients from the resolved fluctuations often generate negative coefficients and instabilities, introducing the need for \textit{ad hoc} clipping or averaging.
One robust, physically-motivated stabilization is the use of Lagrangian averaging \cite{Meneveau1996}.
Other options exist for stabilizing dynamic models, such as choosing a model form based on the full velocity gradient tensor rather than the strain-rate tensor \cite{Rozema2022}. 
In that case, the model does not conform to the need for a symmetric tensor to describe the residual stress, with side-effects such as the loss of global conservation of angular momentum \cite{Aris1962, Batchelor1967}.

Given the lack of timescale separation between the smallest resolved scales and the largest unresolved scales, it is expected that memory effects could be important for an accurate representation \cite{Ballouz2020}. 
The Lagrangian-averaged dynamic procedure \cite{Meneveau1996} encodes some history, but other options exist. 
Li et al. \cite{Li2009} introduced a matrix exponential stress closure based on recent Lagrangian history, and Johnson \cite{Johnson2020b} demonstrated that the nonlinear gradient model naturally encapsulates some important aspects of Lagrangian physics.
Flow history effects can also be incorporated by solving additional transport equations, most commonly for the residual kinetic energy \cite{Schumann1975, Ghosal1995, Kim1995, Ranjan2020}.

A more systematic approach to LES modeling can consider the ideal properties that a residual stress model should fulfill \cite{Silvis2017}.
This often can lead to more elaborate eddy viscosity models \cite{Vreman2004, Rozema2015}, but also models with properties similar to the nonlinear gradient model \cite{Silvis2019}.
Indeed, from the perspective of continuum mechanics, a tensorial approach can be useful to identify a set of coefficients that can be tuned by a traditional dynamic procedure \cite{Agrawal2022}, or more novel ones \cite{Stallcup2022}.
Alternatively, machine learning techniques such as neural networks could be deployed to learn coefficients for tensor-basis expansions \cite{Ling2016}. 
Desirable invariance properties could also be encoded into neural-network based models by training in the strain-rate eigenframe \cite{Prakash2022}.
The future development of data-driven LES models depends, in part, on the effectiveness of the spatial filtering framework itself, or potential alternatives \cite{Geurts2003, Geurts2006, Pruett2008, Pope2010, Fox2012}, including machine learning approaches that altogether circumvent the use of explicit filtering for sub-grid stress modeling \cite{Lozano2023}.

The spatially-filtered Navier-Stokes equation commonly cited as a basis for LES is not without its shortcomings, however.
For example, it is not clear how to conceive of a filter in the near-wall region, where a standard filtering operation would require information outside of the domain \cite{Bhattacharya2008}.
More generally, filters blur out discontinuities (e.g. two-phase interfaces) and generate `spurious parasitic contributions' \cite{Sagaut2005} that complicate the modeling task when specific numerical methods may instead fully-resolve the sharp interface (e.g., sharp interface methods for two-phase flows), albeit with smaller-scale curvature features unresolved.
A more widely recognized difficulty with spatial filtering is the lack of commutativity between filtering and spatial differentiation with the filter size is spatially non-uniform \cite{Ghosal1995Basic, Langford2001, Moser2021}.
This appears to be a relatively practical concern, given the need for non-uniform computational grids for most practical flow simulations.
Indeed, there as been some effort to treat commutator effects \cite{Yalla2021, Kamble2022}, but such commutator errors are generally neglected for convenience in practice.

Johnson \cite{Johnson2022Alternative} proposed an extension of (integral-based) spatial filtering theory based on a differential representation of flow coarsening.
In fact, Germano briefly introduced (elliptic) differential filters decades ago \cite{Germano1986Elliptic, Germano1986LES}.
The physics-inspired approach of Johnson \cite{Johnson2022Alternative} introduced the use of the Stokes equation as a means of artificially coarsening turbulent flows. This Stokes Flow Regularization (SFR) technique is equivalent to spatial filtering with a Gaussian kernel for single-phase unbounded flows with uniform resolution.
However, SFR provides a promising approach to generalize filter-based LES theory to more complex flows.
For example, SFR provides governing equations for a non-uniformly coarsened velocity field that are free of commutator errors.
This paper expands on the preliminary ideas of Ref.\ \cite{Johnson2022Alternative} and uses SFR to generate a range of LES models that are subsequently tested, with an eye toward future extension to more complex flows.
For example, Ref.\ \cite{chen2021interface} introduced the use of diffusion equations to coarsen multi-phase flows while preserving sharp interfaces.

In this paper, the interaction between residual stress models and the energy cascade near the resolution scale is studied by considering the extent of an artificial bottleneck effect and associated spectral bump in LES predictions. To facilitate a fair comparison, the Stokes Flow Regularization (SFR) is extended beyond the results of Ref.\ \cite{Johnson2022Alternative} as follows.
Ref.\ \cite{Johnson2022Alternative} utilized (global) spatial averaging over homogeneous directions to stabilize SFR-based dynamic coefficients. In this paper, it will be shown that practical stabilization techniques arise from SFR theory itself, without the need to invent any \textit{ad hoc} strategies. It will be shown that this theory-based stabilization is equally applicable to models which depart from the eddy viscosity assumption, such as mixed models. Also, SFR theory is herein extended to produce one-equation LES stress models based on a transport equation for residual kinetic energy. The one-equation models provide the ability to capture basic Lagrangian history effects and energetic flow physics that can be particularly useful for future extentions to more complex flows. The result of these extensions is a deeper dive into SFR theory and its implications for LES modeling with an eye towards addressing the artificial bottleneck effect by connecting it to kinetic energy dynamics near the resolution scale.


A preliminary demonstration of the artificial bottleneck effect is given in Section \ref{sec:bottleneck}. Then, Section \ref{sec:theory} reviews and extends the Stokes Flow Regularization (SFR) theory with detailed attention to resolved and residual kinetic energy. Section \ref{sec:SFR models} uses SFR theory to derive a range of models to facilitate a fair comparison of how the form of a model effects the artificial bottleneck effect. Results of LES of isotropic turbulence are shown in Section \ref{sec:results}, and conclusions are drawn in Section \ref{sec:conclusions}.

\section{The artificial Bottleneck effect in LES}\label{sec:bottleneck}

\begin{figure}[htb]
\begin{minipage}{0.5\textwidth}
  \begin{tikzpicture}
  \node (img)  {\includegraphics[scale=1]{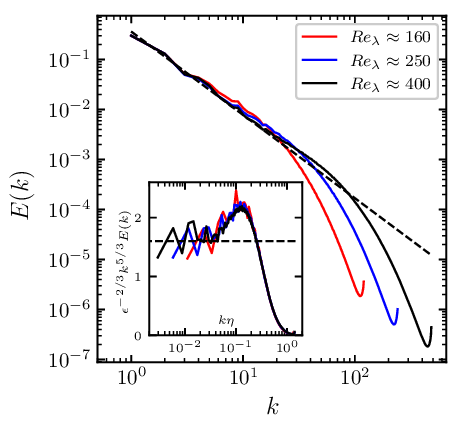}};
    \node[above=of img, node distance=0cm, xshift=-3.5cm, yshift=-1.5cm,font=\color{black}] {(a)};
  \end{tikzpicture}
\end{minipage}%
\begin{minipage}{0.5\textwidth}
  \begin{tikzpicture}
  \node (img)  {\includegraphics[scale=1]{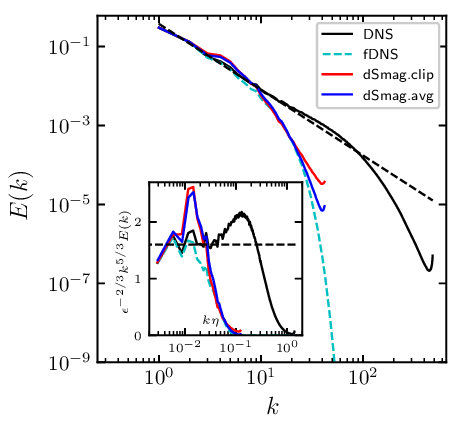}};
  \node[above=of img, node distance=0cm, xshift=-3.5cm, yshift=-1.5cm,font=\color{black}] {(b)};
  \end{tikzpicture}
\end{minipage}%
\caption{ Energy spectra from homogeneous isotropic turbulence for: (a) DNS at different $Re_{\lambda}$, and (b) LES dynamic Smagrinsky (dSmag) with clipping (dSmag.clip) and with global averaging (dSmag.avg) compared to filtered DNS  with $\ell=24 \eta$ at $\kappa_{max} \ell=3.0$. A Kolmogorov
spectrum, $E(\kappa)=1.6 \; \varepsilon^{\frac{2}{3}} \kappa^{-\frac{5}{3}} $ is shown for reference in both panels. The inset in panel (a) and (b) shows the premultiplied spectrum on
a log–linear plot.}
\label{fig:bottleneck}
\end{figure}

Figure \ref{fig:bottleneck}(a) demonstrates the spectral bump observed in DNS of homogeneous isotropic turbulence (HIT) at three different Reynolds numbers, shown most visibly in the premultiplied spectrum plotted on a linear scale (see inset). The spectral bump occurs at a fixed range of wavenumbers normalized by the Kolmogorov length scale. For comparison, energy spectra from LES using a dynamic Smagorinsky (dSmag) model \cite{Germano1991, Lilly1992} are shown in Figure \ref{fig:bottleneck}(b). Here, a Gaussian test filter is used with a test filter scale $2\ell$, which supplies the proper amount of energy dissipation to cause the resulting energy spectra to drop off in an accurate manner near the resolution scale. A similar spectral bump is observed at scales slightly larger than the filter scale.  The spectral bump is observed both for simulations in which the coefficient is clipped (dSmag.clip) and when the coefficient is globally averaged (dSmag.avg). The artificial overshoot in the LES spectra results in an over-prediction of the resolved kinetic energy compared to filtered DNS (fDNS) by about $10\%$, despite the fact that the dynamic procedure is successful in choosing a model coefficient leading to accurate energy removal and spectral roll off at the right wavenumber. That is, the shape of the spectrum is not predicted correctly due to the artificial bottleneck effect which can be expected for any eddy viscosity model. The dynamic Smagorinsky model alone does not predict the residual kinetic energy, but could be complimented with a dynamic Yoshizawa model \cite{yoshizawa1985statistically, moin1991dynamic} to accomplish it. This is not typically done for incompressible flows.

The existence of an artificial bottleneck effect in LES with the dynamic Smagorinski model is not surprising, because it is similar to Navier-Stokes turbulence in its use of a Newtonian viscous stress constitutive law with the local stress tensor proportional to the local strain-rate tensor. This is expected to be a feature of eddy viscosity models more generally. The spectral bump in Navier-Stokes turbulence at high Reynolds number makes a negligible contribution to the total kinetic energy, because the overshoot is associated with length scales near the viscous cutoff. In LES, however, the grid or filter cutoff is generally much closer to the energetic range of scales, so that the artificial bottleneck effect demonstrated in Figure \ref{fig:bottleneck} can be important for simulating high Reynolds number flows. This motivates further investigation into understanding how the form of the residual stress model, grid resolution, filter scale and Reynolds number affect this behaviour. Our paper focuses on analyzing the energy spectrum, turbulent kinetic energy, dissipation, and related statistical parameters such as skewness and flatness. To understand the ability of LES models to accruately represent the details of the energy cascade, the grid scale flow structure is explored by looking at the efficiency of vortex stretching and strain rate self amplification \cite{Johnson2021Role}.

\section{Background \& Theory}
\label{sec:theory}


This section outline the theoretical framework that will guide our investigation into the artificial bottleneck effect. Our focus centers on using the Stokes Flow Regularization (SFR), which serves as a foundational tool for constructing a range of residual stress models using a consistent dynamic procedure that does not require the use of a test filter. First, the background theory of SFR from previous work \cite{Johnson2022Alternative}, followed by a detailed examination of kinetic energy and the cascade through the lens of SFR that extends the results of Ref.\ \cite{Johnson2022Alternative}. Then the SFR-based dynamic procedure for (LES) modeling is introduced.

\subsection{Stokes Flow Regularization: the Basics}

\begin{figure}[h]
    \centering
    \includegraphics[scale=1]{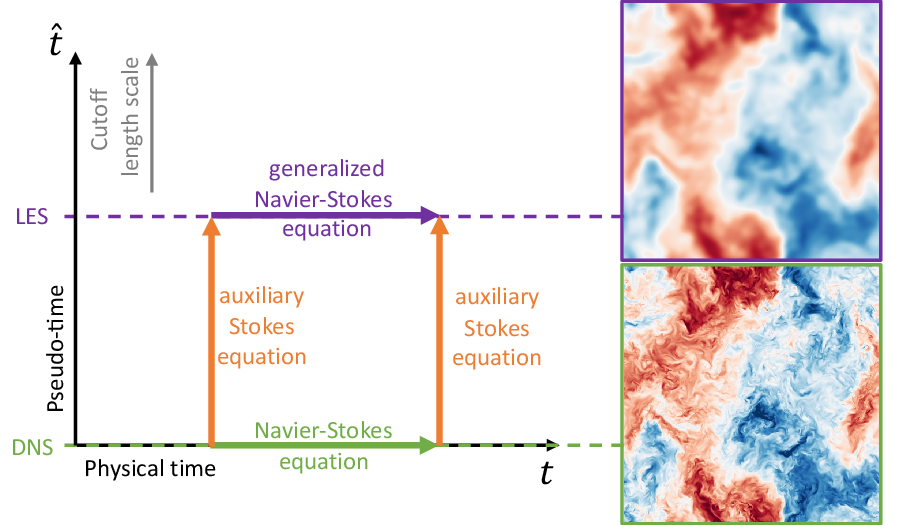}
    \caption{Stokes flow regularization (SFR) for LES defines a generalized (resolved) velocity field as a function of physical time $(t)$ and pseudo-time $(\hat{t})$. The generalized Navier–Stokes equation for the resolved velocity in physical time requires a closure model.}
    \label{fig:SFR-schematic}
\end{figure}

The main concept of Stokes flow regularization (SFR) is sketched in Figure \ref{fig:SFR-schematic}. A resolved velocity field, 
$\mathbf{U}(\mathbf{x}, t, \hat{t})$, is introduced as a function of both physical time, $t$, and a new time-like variable that measures the extent to which small-scale features are removed from the representation of the flow: the pseudo-time, $\hat{t}$.

When $\hat{t} = 0$, the resolved velocity field must be equivalent to the physical velocity field,
\begin{equation}
    \mathbf{U}(\mathbf{x}, t; 0)
    = \mathbf{u}(\mathbf{x}, t).
    \label{eq:aux-IC-velocity}
\end{equation}
This grounds the basis of SFR in physical reality. The physical velocity field obeys the fundamental laws of mass, momentum, and energy conservation. For incompressible Newtonian flows,
\begin{equation}
    \frac{\partial u_i}{\partial t} + u_j \frac{\partial u_i}{\partial x_j} = - \frac{\partial p}{\partial x_i} + \nu \nabla^2 u_i + f_i,
    \hspace{0.1\linewidth}
    \frac{\partial u_j}{\partial x_j} = 0,
    \label{eq:Navier-Stokes}
\end{equation}
where $p(\mathbf{x}, t)$ is the pressure (divided by density, $\rho$) and $\mathbf{f}(\mathbf{x}, t)$ denotes any additional solenoidal body forces, e.g., from gravity or artificial stirring. The kinematic viscosity is $\nu = \mu/\rho$. For incompressible flows, the pressure is responsible for enforcing the divergence-free condition during the evolution of the velocity field in (physical) time. Using the divergence of Eq. \eqref{eq:Navier-Stokes}, the pressure may be found as the solution of a Poisson equation,
\begin{equation}
    \nabla^2 p = - \frac{\partial u_i}{\partial x_j} \frac{\partial u_j}{\partial x_i}. 
    \label{eq:pressure-Poisson}
\end{equation}
Equations \eqref{eq:Navier-Stokes} and \eqref{eq:pressure-Poisson} are considered sufficient to describe the observed velocity field, $\mathbf{u}(\mathbf{x}, t)$, of low-Mach turbulent flows. The implied evolution of kinetic energy is,
\begin{equation}
    \frac{\partial}{\partial t} \left(\frac{1}{2}u_iu_i \right)
    + \frac{\partial }{\partial x_j} \left(
        \frac{1}{2}u_i u_i u_j + u_i p - 2 \nu u_i s_{ij}
    \right)
    =
    - 2 \nu s_{ij} s_{ij}
    + u_i f_i
    \label{eq:kinetic-energy-transport}
\end{equation}
where $s_{ij} = \tfrac{1}{2} \left( \frac{\partial u_i}{\partial x_j} + \frac{\partial u_j}{\partial x_i} \right)$ is the rate-of-strain tensor.

In practice, however, it is desirable to have the capability to solve for the dynamics of a coarse-grained representation of the flow. Dating back to Ref.\ \cite{Leonard1975}, the basis of LES has been conceived as the solution for the spatially-filtered velocity field,
\begin{equation}
    \overline{u}_i^\ell(\mathbf{x}, t)
    =
    \iiint G_\ell(\mathbf{r}) u_i(\mathbf{x}+\mathbf{r}, t) d\mathbf{r}
    \label{eq:filtered-velocity}
\end{equation}
with a suitable normalized filter kernal, $G$. The SFR concept generalizes the spatial filtering approach. More generally, a turbulent flow snapshot may be coarsened by deploying the physical equations with the energy cascade mechanism removed, namely, the nonlinear inertia term in Eq.\ \eqref{eq:Navier-Stokes}, $u_j ~\partial u_i/ \partial x_j$. The result is the Stokes equation, which is physically valid only for low Reynolds number flows, but is useful here in a different way: as an artificial coarsening procedure. Thus, the Stokes equation describes the evolution of the generalized velocity in pseudo-time (at a fixed physical time),
\begin{equation}
    \frac{\partial U_i}{\partial \hat{t}} = - \frac{\partial \hat{p}}{\partial x_i} + \frac{\partial}{\partial x_j}\left( 2 \hat{\nu} S_{ij} \right),
    \hspace{0.1\linewidth}
    \frac{\partial U_j}{\partial x_j} = 0,
    \label{eq:aux-Stokes}
\end{equation}
where $S_{ij} =\tfrac{1}{2} \left( \frac{\partial U_i}{\partial x_j} + \frac{\partial U_j}{\partial x_i} \right)$ is the generalized strain-rate tensor. The orange lines in Fig.~\ref{fig:SFR-schematic} represent Eq.\ \eqref{eq:aux-Stokes}. A pseudo-viscosity, $\hat{\nu}$, introduced to accomplish the coarsening during pseudo-time, may be made into a anisotropic tensorial pseudo-viscosity for anisotropic grid resolution and may be made a function of space to accomodate spatially non-uniform grid resolution. 
Note that the pseudo-viscosity is an element of the coarsening procedure that generalized spatial filtering. 
It is different than the physical viscosity, and it is not the same as an eddy viscosity commonly used in turbulence modeling.
The pseudo-pressure, $\hat{p}$, is responsible for enforcing the divergence free condition during pseudo-time evolution, and may be found by solution of a Poisson equation resulting from the divergence of Eq.\ \eqref{eq:aux-Stokes}. 
The local resolution of the generalized velocity field at pseudo-time $\hat{t}$ may be estimates as $\ell \sim \sqrt{\hat{\nu} \hat{t}}$.

According to Eq.\ \eqref{eq:aux-IC-velocity}, the dynamics of the resolved velocity at zero pseudo-time, $\mathbf{U}(\mathbf{x}, t; 0)$, must adhere to the incompressible Navier-Stokes dynamics of Eqs.\ \eqref{eq:Navier-Stokes} and \eqref{eq:pressure-Poisson} during their evolution in physical time for $\hat{t} = 0$. The corresponds to the green arrow in Fig.\ \ref{fig:SFR-schematic}. However, the physical time evolution at a non-zero pseudo-time must inevitably differ from the Navier-Stokes equation. Thus, a residual force, $\boldsymbol{\Phi}(\mathbf{x}, t; \hat{t})$, is introduced to create a generalized Navier-Stokes equation capable of describing resolved flow dynamics at a finite pseudo-time,
\begin{equation}
    \frac{\partial U_i}{\partial t}
    + U_j \frac{\partial U_i}{\partial x_j}
    =
    - \frac{\partial P}{\partial x_i}
    + \nu \nabla^2 U_i
    + F_i
    + \Phi_i,
    \hspace{0.1\linewidth}
    \frac{\partial U_j}{\partial x_j} = 0.
    \label{eq:gen-Navier-Stokes-1}
\end{equation}
Equation \eqref{eq:gen-Navier-Stokes-1} represents the effective evolution of a coarse-grained representation of the velocity field, and must reduce to the Navier-Stokes equation, Eq.\ \eqref{eq:Navier-Stokes}, in the limit $\hat{t} \rightarrow 0$.
The generalized (resolved) pressure, $P(\mathbf{x}, t; \hat{t})$, enforces the divergence-free condition during physical time evolution at finite pseudo-time; it can be found using a Poisson equation generated by taking the divergence of Eq.\ \eqref{eq:gen-Navier-Stokes-1}. Therefore, the reduction of Eq.\ \eqref{eq:gen-Navier-Stokes-1} to Eq.\ \eqref{eq:Navier-Stokes} in the limit $\hat{t} \rightarrow 0$ requires,
\begin{equation}
    P(\mathbf{x}, t; 0) = p(\mathbf{x}, t).
    \label{eq:aux-IC-pressure}
\end{equation}
The generalized body force, $\mathbf{F}(\mathbf{x}, t; \hat{t})$, likewise represents the effective body force at coarse resolution, thus,
\begin{equation}
    F_i(\mathbf{x}, t; 0) = f_i(\mathbf{x}, t).
    \label{eq:aux-IC-body-force}
\end{equation}
Finally, the residual force, $\boldsymbol{\Phi}(\mathbf{x}, t; \hat{t})$, represents the residual error of the Navier-Stokes equation applied at coarse resolution (finite pseudo-time), therefore,
\begin{equation}
    \Phi_i(\mathbf{x}, t; 0) = 0.
    \label{eq:aux-IC-residual-force}
\end{equation}
In general, this residual force may be specified by enforcing the equality of mixed partial derivatives in the $t$-$\hat{t}$ phase space shown in Fig. \ref{fig:SFR-schematic},
\begin{equation}
    \frac{\partial}{\partial t}\left( \frac{\partial U_i}{\partial \hat{t}} \right)
    =
    \frac{\partial}{\partial \hat{t}}\left( \frac{\partial U_i}{\partial t} \right).
    \label{eq:mixed-partials-velocity}
\end{equation}

\subsection{A Special Case: Uniform Resolution of Unbounded Flows}
For demonstration purposes, let us proceed with the derivation assuming uniform, isotropic resolution of a flow on an unbounded domain. Therefore, the Stokes equation, Eq.\ \eqref{eq:aux-Stokes} with initial condition given by Eq.\ \eqref{eq:aux-IC-velocity}, simplifies to a diffusion equation,
\begin{equation}
    \frac{\partial U_i}{\partial \hat{t}} = \hat{\nu} \nabla^2 U_i,
    \hspace{0.1\linewidth}
    U_i(\mathbf{x}, t; 0) = u_i(\mathbf{x}, t). 
    \label{eq:aux-diffusion-velocity}
\end{equation}
In this case, the solution to the diffusion equation is straightforward,
\begin{equation}
    U_i(\mathbf{x}, t; \hat{t})
    =
    \iiint G(\mathbf{r}; \hat{t}) u_i(\mathbf{x}+\mathbf{r}, t) d\mathbf{r},
    \hspace{0.02\linewidth}
    \text{where}
    \hspace{0.02\linewidth}
    G(\mathbf{r}; \hat{t}) = \frac{1}{\left( 4 \pi \hat{\nu} \hat{t} \right)^{3/2}} \exp\left( -\frac{|\mathbf{r}|^2}{4 \hat{\nu} \hat{t}} \right),
    \label{eq:aux-diffusion-velocity-solution}
\end{equation}
which is equivalent to a Gaussian spatial filter with $\ell^2 = 2 \hat{\nu} \hat{t}$. This establishes a formal mathematical equivalence between SFR and spatial filtering in the special case of unbounded flows with uniform resolution.

Equations \eqref{eq:aux-diffusion-velocity} and \eqref{eq:gen-Navier-Stokes-1} may be substituted in Eq. \eqref{eq:mixed-partials-velocity} on the left and right hand side, respectively, and further simplifications and substitutions  lead to,
\begin{equation}
    \frac{\partial \Phi_i}{\partial \hat{t}} = 
    \hat{\nu} \nabla^2 \Phi_i
    + \frac{\partial}{\partial x_j}\left[ 
        - 2 \hat{\nu} A_{ik} A_{jk}
        + \left( \frac{\partial P}{\partial \hat{t}} - \hat{\nu} \nabla^2 P \right) \delta_{ij} 
    \right]
    - \left( \frac{\partial F_i}{\partial \hat{t}} - \hat{\nu} \nabla^2 F_i \right),
    \label{eq:aux-diffusion-residual-force}
\end{equation}
where $A_{ij}(\mathbf{x}, t; \hat{t}) = \tfrac{\partial U_i}{\partial x_j}$ is the resolved velocity gradient tensor at pseudo-time $\hat{t}$. It is most natural to separate the body force from the residual force by choosing,
\begin{equation}
    \frac{\partial F_i}{\partial \hat{t}} = \hat{\nu} \nabla^2 F_i,
    \hspace{0.1\linewidth}
    F_i(\mathbf{x}, t; 0) = f_i(\mathbf{x}, t)
    \label{eq:aux-diffusion-body-force}
\end{equation}
where the initial condition comes from Eq.\ \eqref{eq:aux-IC-body-force}. Similarly, the generalized (resolved) pressure can absorb the isotropic part of the velocity gradient product to form,
\begin{equation}
    \frac{\partial P}{\partial \hat{t}} = \hat{\nu} \nabla^2 P + \frac{2}{3} \hat{\nu} A_{ik} A_{ik},
    \hspace{0.1\linewidth}
    P(\mathbf{x}, t; 0) = p(\mathbf{x}, t)
    \label{eq:aux-diffusion-pressure}
\end{equation}
This equation is a linear force diffusion equation, and the solution can be written via superposition as $P = P^{(1)} + P^{(2)}$ where
\begin{equation}
    \frac{\partial P^{(1)}}{\partial \hat{t}} = \hat{\nu} \nabla^2 P^{(1)},
    \hspace{0.1\linewidth}
    P^{(1)}(\mathbf{x}, t; 0) = p(\mathbf{x}, t),
    \label{eq:ge-pressure-1}
\end{equation}
\begin{equation}
    \frac{\partial P^{(2)}}{\partial \hat{t}} = \hat{\nu} \nabla^2 P^{(2)}
    + \frac{2}{3} \hat{\nu} A_{ik} A_{ik},
    \hspace{0.1\linewidth}
    P^{(2)}(\mathbf{x}, t; 0) = 0,
    \label{eq:gen-pressure-2}
\end{equation}
The first component, $P^{(1)}$, is the Gaussian-filtered pressure while the second component, $P^{(2)}$, relates to the portion of residual kinetic energy that may be lumped into the effective pressure for incompressible flows.

With this choice for the generalized (resolved) body force and pressure, the remaining source term in the forced diffusion equation for the residual force, Eq.\ \eqref{eq:aux-diffusion-residual-force}, is inside of a divergence. It follows that the residual force must be a divergence of a residual stress,
\begin{equation}
    \Phi_i = -\frac{\partial \sigma_{ij}}{\partial x_j}
    \label{eq:residual-force-as-stress-divergence}
\end{equation}
Physically, this must be true because the Stokes equation globally conserves momentum during pseudo-time evolution, so the residual force it gives rise to must likewise globally conserve momentum. The result is a generalized Navier-Stokes equation for the resolved velocity dynamics,
\begin{equation}
    \frac{\partial U_i}{\partial t} 
    + U_j \frac{\partial U_i}{\partial x_j} 
    = 
    - \frac{\partial P}{\partial x_i} 
    + \nu \nabla^2 U_i
    - \frac{\partial \sigma_{ij}}{\partial x_j} 
    + F_i,
    \hspace{0.1\linewidth}
    \frac{\partial U_j}{\partial x_j} = 0,
    \label{eq:gen-Navier-Stokes}
\end{equation}
where $\boldsymbol{F}$ represents the coarsened body force and the resolved pressure, $P$, may be found by enforcing the divergence-free condition using a Poisson equation. Enforcing Eqs. \eqref{eq:aux-diffusion-body-force}, \eqref{eq:aux-diffusion-pressure}, and \eqref{eq:residual-force-as-stress-divergence}, then Eq.\ \eqref{eq:aux-diffusion-residual-force} becomes, 
\begin{equation}
    \frac{\partial}{\partial x_j} \left( \frac{\partial \sigma_{ij}}{\partial \hat{t}}
    - \hat{\nu} \nabla^2 \sigma_{ij} 
    - 2 \hat{\nu} (A_{ik} A_{jk})^{(d)} \right)
    = 0,
    \hspace{0.1\linewidth}
    \frac{\partial \sigma_{ij}}{\partial x_j}(\mathbf{x}, t; 0) = 0,
    \label{eq:aux-diffusion-residual-stress-div}
\end{equation}
where the superscript $^{(d)}$ denotes the deviatoric component of a tensor.
While any $\sigma_{ij}$ satisfying Eq.\ \eqref{eq:aux-diffusion-residual-stress-div} is sufficient, it can simplify the modeling process to consider a more strict criterion, namely, that the residual stress tensor is defined as the solution to a forced diffusion equation,
\begin{equation}
    \frac{\partial \sigma_{ij}}{\partial \hat{t}}
    = \hat{\nu} \nabla^2 \sigma_{ij} 
    + 2 \hat{\nu} (A_{ik} A_{jk})^{(d)},
    \hspace{0.1\linewidth}
    \sigma_{ij}(\mathbf{x}, t; 0) = 0.
    \label{eq:aux-diffusion-residual-stress}
\end{equation}
The ability to add (or subtract) divergence-free tensor fields to the residual stress tensor has been discussed in detail by Ref.\ \cite{VelaMartin2022}. 
 The formal solution for unbounded flow with uniform resolution is,
\begin{equation}
    \sigma_{ij}(\mathbf{x}, t; \hat{t}) = 2 \hat{\nu}
    \int_0^{\hat{t}}
    \left( \iiint (A_{ik} A_{jk})^{(d)}(\mathbf{x}+\mathbf{r}, t; \hat{t}^\prime) G(\mathbf{r}, \hat{t}-\hat{t}^\prime) d\mathbf{r} \right)
    d\hat{t}^\prime,
    \label{eq:residual-stress-solution}
\end{equation}
where the function $G(\mathbf{r}, \hat{t})$ is the same as used in Eq.\ \eqref{eq:aux-diffusion-velocity-solution}. Thus, the residual stress at scale $\ell = \sqrt{2 \hat{\nu} \hat{t}}$ arises from velocity gradients at all scales $0 \leq \ell^\prime \leq \ell$.

\subsection{Kinetic Energy in Stokes Flow Regularization}

Turbulent flows are famous for their rapid mixing capability, driven by an enhanced ability to dissipate kinetic energy. 
The enhanced dissipation in turbulence is understood as the outcome of a chaotic cascading processes in which small-scale fluctuations are rapidly energized from large-scale excitation by the inherent nonlinearity of high Reynolds number fluid dynamics.
The kinetic energy cascade is one of the hallmark concepts in turbulence theory, dating at least back to Richardson's \cite{Richardson1922} rhyming description of the interaction between ``big whirls,'' ``little whirls,'' and ``lesser whirls.'' 
In a footnote of Kolmogorov's seminal paper introducing his hypotheses about locally homogeneous and isotropic turbulence, a similar description is given in terms of a hierarchy of ``pulsations'' \cite{Kolmogorov1941a}. 
Around the same time, Taylor studied the dynamics of turbulence generated behind a grid in a wind tunnel and concluded that ``the stretching of vortex filaments must be regarded as the principal mechanical cause of the high rate of dissipation which is associated with turbulent motion'' \cite{Taylor1938}.
Onsager \cite{Onsager1949} coined the terminology of ``cascade'' \cite{Eyink2006Onsager} and tied it to the vortex stretching mechanism of Taylor.
Thus, a common understanding that emerged from these early works on turbulence theory was that the energy cascade is driven by multiscale vortex stretching \cite{Tennekes1972}.

More recent studies have highlighted a second cascade mechanism at work: strain-rate self-amplification \cite{Tsinober2009Informal}. Carbone \& Bragg \cite{Carbone2020Vortex} used the dynamical equations for velocity increments to estimate that strain-rate self-amplification is responsible for $75\%$ of the energy cascade rate while vortex stretching is only responsible for the remaining $25\%$. Similar estimates can be obtained using spatial filtering \cite{Eyink2006Multiscale}. It is encouraging to observe the fundamental agreement between increment-based and filtering-based estimates of how much vortex stretching and strain-rate self-amplification contribute to the the energy cascade rate. This lends some evidence that these conclusions are features of the underlying physics, independent of the particulars of the analysis method. However, the $75\%$:$25\%$ partition between strain-rate self-amplification and vortex stretching contributions to the energy cascade rate relies on the truncation of an infinite series expansion, regardless of the analysis method, which leaves uncertainty as to the role of higher-order terms in such an approximation. This is precisely the point at which SFR provides a pivotal breakthrough.

Although SFR was introduced for LES as a ``physics-inspired coarsening'' technique in Ref. \cite{Johnson2022Alternative}, the use of similar techniques for theoretical fluid mechanics predates this recent work. In unpublished notes, Onsager appears to have considered Stokes flow as a natural basis for analysis and modeling of turbulent channel flow \cite{Eyink2006Onsager}. Isett and Oh \cite{Isett2016Heat} used a ``geometric heat flow'' regularization for proofs about the nature of weak Euler solutions on a more general class of Riemannian manifolds. Johnson \cite{Johnson2020Energy, Johnson2021Role} demonstrated the use of SFR for analyzing the physics of the energy cascade in terms of vortex stretching and strain-rate self-amplification, which is summarized here before returning to the application of SFR to LES modeling. To do so, the transport equations for kinetic energy within the SFR framework are developed.


The Stokes flow regularization, Eq.\ \eqref{eq:aux-diffusion-velocity}, removes kinetic energy as pseudo-time increases,
\begin{equation}
    \frac{\partial E}{\partial \hat{t}} = \hat{\nu} \nabla^2 E - \hat{\nu} A_{ij} A_{ij},
    \hspace{0.1\linewidth}
    E(\mathbf{x}, t; \hat{t}=0) = \frac{1}{2} u_i u_i (\mathbf{x}, t),
    \label{eq:aux-diffusion-resolved-energy}
\end{equation}
where $E = \tfrac{1}{2} U_i U_i$ is the kinetic energy of the resolved velocity field at any pseudo-time. It represents the kinetic energy of fluid motions larger than the resolution scale $\ell = \sqrt{2 \hat{\nu} \hat{t}}$. From observation of Eq.\ \eqref{eq:gen-pressure-2}, a residual (or small-scale) kinetic energy can be defined as $e = \frac{3}{2}P^{(2)}$, so that,
\begin{equation}
    \frac{\partial e}{\partial \hat{t}} = \hat{\nu} \nabla^2 e + \hat{\nu} A_{ij} A_{ij},
    \hspace{0.1\linewidth}
    e(\mathbf{x}, t; \hat{t}=0) = 0,
    \label{eq:aux-diffusion-residual-energy}
\end{equation}
such that the sum of these two energies, $T = E + e$, is globally conserved in pseudo-time,
\begin{equation}
    \frac{\partial T}{\partial \hat{t}} = \hat{\nu} \nabla^2 T,
    \hspace{0.1\linewidth}
    T(\mathbf{x}, t; \hat{t}=0) = \frac{1}{2} u_i u_i (\mathbf{x}, t).
    \label{eq:aux-diffusion-total-energy}
\end{equation}
In this way SFR provides a convenient separation of kinetic energy into large- and small-scale components, $E$ and $e$, respectively. Equations \eqref{eq:aux-diffusion-resolved-energy} and \eqref{eq:aux-diffusion-residual-energy} show how the energy is being removed from resolved scales to unresolved scales by the term $\hat{\nu} A_{ij} A_{ij}$ in pseudo time. 

The evolution of large-scale kinetic energy, $E$, in physical time (at fixed pseudo-time) can be directly inferred from the resolved dynamics, Eq.\ \eqref{eq:gen-Navier-Stokes},
\begin{equation}
    \frac{\partial}{\partial t} \left(\frac{1}{2}U_i U_i \right)
    + \frac{\partial}{\partial x_j} \left(
        \frac{1}{2}U_i U_i U_j 
        + U_j P
        - 2 \nu U_i S_{ij}
        + U_i \sigma_{ij}
    \right)
    =
    \sigma_{ij} S_{ij}
    - 2 \nu S_{ij} S_{ij}
    + U_i F_i
    \label{eq:resolved-kinetic-energy-transport}
\end{equation}
The evolution of total energy, $T = E + e$, in physical time may be found using a similar procedure as for the momentum. The total kinetic energy must equal the physical kinetic energy when $\hat{t} = 0$, so its evolution equation can be written as the physical kinetic energy transport equation, Eq.\ \eqref{eq:kinetic-energy-transport}, plus a to-be-determined residual error ($W$), 
\begin{equation}
    \frac{\partial T}{\partial t}
    + \frac{\partial}{\partial x_j} \left(
        \frac{1}{2}U_i U_i U_j + U_j P^{(1)} - 2 \nu U_i S_{ij}
    \right)
    =
    - 2 \nu S_{ij} S_{ij}
    + U_i F_i
    + W.
    \label{eq:total-kinetic-energy-transport-1}
\end{equation}
The residual work, $W$, can be specified by requiring the equality of mixed partial derivatives, following Eq.\ \eqref{eq:mixed-partials-velocity},
\begin{equation}
    \frac{\partial}{\partial t}\left(\frac{\partial T}{\partial \hat{t}} \right )
    =
    \frac{\partial}{\partial \hat{t}}\left(\frac{\partial T}{\partial t} \right ).
    \label{eq:mixed-partials-energy}
\end{equation}
After substitution of Eqs.\ \eqref{eq:aux-diffusion-total-energy} and \eqref{eq:total-kinetic-energy-transport-1} into Eq.\ \eqref{eq:mixed-partials-energy}, with further manipulation, a superposition solution for the residual work may be written, where
\begin{equation}
    {W} = - \frac{\partial}{\partial x_j}\left( q_j^{(1)} + q_j^{(2)} + q_j^{(3)} \right) + {W}^{(4)} + {W}^{(5)}
    \label{eq:residual-work-decomposition}
\end{equation}
\begin{equation}
    \frac{\partial q_j^{(1)}}{\partial \hat{t}}
    =
    \hat{\nu} \nabla^2 q_j^{(1)}
    + 2 \hat{\nu} \frac{\partial U_i}{\partial x_k} \frac{\partial \sigma_{ij}}{\partial x_k}
    + 2 \hat{\nu} \frac{\partial U_j}{\partial x_k} \frac{\partial e}{\partial x_k}
    \label{eq:residual flux-1}
\end{equation}
\begin{equation}
    \frac{\partial q_j^{(2)}}{\partial \hat{t}}
    =
    \hat{\nu} \nabla^2 q_j^{(2)}
    + 2\hat{\nu} \frac{\partial U_j}{\partial x_k} \frac{\partial P}{\partial x_k}
    \label{eq:residual flux-2}
\end{equation}
\begin{equation}
    q_j^{(3)} = - \nu \frac{\partial e}{\partial x_j}
\end{equation}
\begin{equation}
    \frac{\partial {W}^{(4)}}{\partial \hat{t}}
    =
    \hat{\nu} \nabla^2 {W}^{(4)}
    - 2 \nu \hat{\nu} \frac{\partial A_{ij}}{\partial x_k} \frac{\partial A_{ij}}{\partial x_k}
\end{equation}
\begin{equation}
    \frac{\partial {W}^{(5)}}{\partial \hat{t}}
    =
    \hat{\nu} \nabla^2 {W}^{(5)}
    + 2 \hat{\nu} \frac{\partial U_{i}}{\partial x_k} \frac{\partial F_{i}}{\partial x_k}
\end{equation}
The first residual flux, $\mathbf{q}^{(1)}$, represents the flux of residual kinetic energy by unresolved turbulent velocity fluctuations. The second residual flux, $\mathbf{q}^{(2)}$, represents the flux of residual kinetic energy by unresolved pressure-velocity coupling. The third flux, $\mathbf{q}^{(3)}$, is the viscous diffusion of residual kinetic energy (this term is closed and requires no modeling). The first residual work, ${W}^{(4)}$, represents viscous dissipation of residual kinetic energy. Finally, the remaining residual work term, ${W}^{(5)}$, includes the direct production of residual kinetic energy by unresolved body forces.

A transport equation for residual kinetic energy may be obtained by subtracting the transport equation for large-scale kinetic energy, Eq.\ \eqref{eq:resolved-kinetic-energy-transport}, from the total kinetic energy transport equation, Eq.\ \eqref{eq:total-kinetic-energy-transport-1}. Then, using the decomposition of residual work, Eq.\ \eqref{eq:residual-work-decomposition}, it is obtained that,
\begin{equation}
    \frac{\partial e}{\partial t}
    + \frac{\partial}{\partial x_j}\left(
        U_j e
        - \nu \frac{\partial e}{\partial x_j}
        + q_j^{(1)}
        + q_j^{(2)}
    \right)
    =
    - \sigma_{ij} S_{ij}
    + {W}^{(4)}
    + {W}^{(5)}
    \label{eq:residual-kinetic-energy-transport}
\end{equation}
Comparing Eq.\ \eqref{eq:residual-kinetic-energy-transport} with Eq.\ \eqref{eq:resolved-kinetic-energy-transport}, it is evident that $\Pi = - \sigma_{ij} S_{ij}$ represents the rate at which large-scale kinetic energy is converted into small-scale kinetic energy. This may be thought of as the rate at which large-scale motions do work on small-scale motions \cite{Ballouz2018Tensor, Tennekes1972}. Invoking Onsager's ``cascade'' imagery, the quantity $\Pi(\mathbf{x}, t; \hat{t})$ represents the kinetic energy cascade rate across scale $\ell = \sqrt{2 \hat{\nu} \hat{t}}$ at any location and time in the flow. 

Using the formal solution for the residual stress tensor, Eq.\ \eqref{eq:residual-stress-solution}, it can be seen that
\begin{equation}
    \Pi(\mathbf{x}, t; \hat{t}) =
    \iiint
    2 \hat{\nu} \left(\int_0^{\hat{t}}
    -A_{ik} (\mathbf{x}+\mathbf{r}, t; \hat{t}^\prime) A_{jk}(\mathbf{x}+\mathbf{r}, t; \hat{t}^\prime) S_{ij}(\mathbf{x}, t; \hat{t})
    d\hat{t}^\prime \right)
    G(\mathbf{r}, \hat{t}-\hat{t}^\prime) d\mathbf{r} 
\end{equation}
Thus, the kinetic energy cascade rate across scale $\ell = \sqrt{2 \hat{\nu} \hat{t}}$ is formally connected to the interaction between the strain-rate at scale $\ell$ and the velocity gradients at all scales $0 \leq \ell^\prime \leq \ell$, where $\ell^\prime = \sqrt{2 \hat{\nu} \hat{t}^\prime}$, within the neighborhood of the larger-scale strain-rate. More specifically, a small-scale velocity gradient may be defined as,
\begin{equation}
    A_{ik}^\prime A_{jk}^\prime (\mathbf{x}, t; \hat{t}; \mathbf{r}, \hat{t}^\prime) =
    A_{ik} A_{jk} (\mathbf{x}+\mathbf{r}, t; \hat{t}^\prime)
    - A_{ik} A_{jk}(\mathbf{x}, t; \hat{t})
\end{equation}
so that $\Pi = \Pi_1 + \Pi_2$, where,
\begin{equation}
    \Pi_1(\mathbf{x}, t; \hat{t}) = - \ell^2 A_{ik} A_{jk} S_{ij} (\mathbf{x}, t; \hat{t}),
    \label{eq:cascade-rate-single-scale}
\end{equation}
\begin{equation}
    \text{and}
    \quad
    \Pi_2(\mathbf{x}, t; \hat{t}) =
    \iiint
    2 \hat{\nu} \left(\int_0^{\hat{t}}
    -A_{ik}^\prime A_{jk}^\prime (\mathbf{x}+\mathbf{r}, t; \hat{t}; \mathbf{r}, \hat{t}^\prime) S_{ij}(\mathbf{x}, t; \hat{t})
    d\hat{t}^\prime \right)
    G(\mathbf{r}, \hat{t}-\hat{t}^\prime) d\mathbf{r}.
    \label{eq:cascade-rate-multi-scale}
\end{equation}
The first term, $\Pi_1$, represents the kinetic energy cascade rate driven by single-scale velocity gradient interactions at scale $\ell = \sqrt{2 \hat{\nu} \hat{t}}$, whereas the second term, $\Pi_2$, represents the kinetic energy cascade rate due to two-scale interactions between the strain-rate at scale $\ell = \sqrt{2 \hat{\nu} \hat{t}}$ and the smaller-scale velocity gradient fluctuations at scale $\ell^\prime = \sqrt{2 \hat{\nu} \hat{t}^\prime}$, where $0 \leq \ell^\prime \leq \ell$.

This result answers the question of how vortex stretching and strain-rate self-amplification relate to the energy cascade. Further decomposing the velocity gradient tensor into its symmetric and anti-symmetric components, $A_{ij} = S_{ij} - \frac{1}{2} \epsilon_{ijk} \Omega_k$, where $\boldsymbol{\Omega}$ is the resolved vorticity vector. Inserting this decomposition, $\Pi_1 = \Pi_{s1} + \Pi_{\omega 1}$ and $\Pi_2 = \Pi_{s2} + \Pi_{\omega 2} + \Pi_{c 2}$, where,
\begin{equation}
    \Pi_{s1} = - \ell^2 S_{ij} S_{jk} S_{ki},
    \hspace{0.1\linewidth}
    \Pi_{\omega 1} = \frac{1}{4} \ell^2 \Omega_i S_{ij} \Omega_j,
    \label{eq:resolved energy cascade}
\end{equation}
\begin{equation}
    \Pi_{s2} = \iiint
    2 \hat{\nu} \left(\int_0^{\hat{t}}
    -S_{ik}^\prime S_{kj}^\prime (\mathbf{x}+\mathbf{r}, t; \hat{t}; \mathbf{r}, \hat{t}^\prime) S_{ij}(\mathbf{x}, t; \hat{t})
    d\hat{t}^\prime \right)
    G(\mathbf{r}, \hat{t}-\hat{t}^\prime) d\mathbf{r},
\end{equation}
\begin{equation}
    \Pi_{\omega 2} = \iiint
    \frac{1}{2} \hat{\nu} \left(\int_0^{\hat{t}}
    \Omega_{i}^\prime \Omega_{j}^\prime (\mathbf{x}+\mathbf{r}, t; \hat{t}; \mathbf{r}, \hat{t}^\prime) S_{ij}(\mathbf{x}, t; \hat{t})
    d\hat{t}^\prime \right)
    G(\mathbf{r}, \hat{t}-\hat{t}^\prime) d\mathbf{r},
\end{equation}
\begin{equation}
    \Pi_{c 2} = \iiint
    2 \hat{\nu} \left(\int_0^{\hat{t}}
    \left( S_{ik}^\prime R_{kj}^\prime - R_{ik}^\prime S_{kj}^\prime\right) (\mathbf{x}+\mathbf{r}, t; \hat{t}; \mathbf{r}, \hat{t}^\prime) S_{ij}(\mathbf{x}, t; \hat{t})
    d\hat{t}^\prime \right)
    G(\mathbf{r}, \hat{t}-\hat{t}^\prime) d\mathbf{r},
\end{equation}
where $R_{ij} = - \frac{1}{2} \epsilon_{ijk} \Omega_k$ is the resolved rotation-rate tensor.

The two single-scale terms, $\Pi_{s1}$ and $\Pi_{\omega 1}$, represent the strain-rate self-amplification and vortex stretching contributions to the cascade rate. The Betchov relation for the third invariant \cite{Betchov1956} of the velocity gradient tensor (at any scale), requires that $\left\langle \Pi_{s1} \right\rangle = 3 \left\langle \Pi_{\omega 1} \right\rangle$, so that $75\%$ of $\langle \Pi_1 \rangle$ may be attributed to strain-rate self-amplification, $\langle \Pi_{s1} \rangle$. The remaining $25\%$ is attributable to $\langle \Pi_{\omega 1} \rangle$. This decomposition of the cascade rate corresponds to that previously obtained by truncation of an infinite series \cite{Eyink2006Multiscale, Carbone2020Vortex}. 

The first two multiscale terms, $\Pi_{s2}$ and $\Pi_{\omega 2}$, represent smaller-scale strain-rate amplification and vortex stretching interactions by larger-scale strain rates. Johnson \cite{Johnson2020Energy, Johnson2021Role} observed that the average values of these two terms were nearly equal in the inertial range of homogeneous isotropic turbulence using direct numerical simulation at $Re_\lambda \approx 400$. Moreover, the final term, $\Pi_{c2}$, represents vortex thinning and its average values was found to be very small compared with the other four terms \cite{Johnson2020Energy, Johnson2021Role}.

The average cascade rate was found to be evenly distributed between $\langle \Pi_1 \rangle$ and $\langle \Pi_2 \rangle$ in the inertial range of scales. Therefore, on the whole, strain-rate amplification, $\langle \Pi_{s1} + \Pi_{s2} \rangle$ accounts for approximately $\tfrac{5}{8}$ of the average energy cascade rate in the inertial range and vortex stretching, $\langle \Pi_{\omega 1} + \Pi_{\omega 2} \rangle$, supplies the remaining $\tfrac{3}{8}$. Considering that $\Pi = \Pi_{s1} + \Pi_{\omega 1} + \Pi_{s2} + \Pi_{\omega 2} + \Pi_{c 2}$ is a point-wise exact relationship for the energy cascade rate, thus SFR provides basic insight into the physics of the energy cascade.

\subsection{Stokes Flow Regularization as a Basis for LES Modeling}
\label{sec:test}
Equation \eqref{eq:gen-Navier-Stokes} describes the physical-time evolution of the resolved velocity field, $\mathbf{U}(\mathbf{x}, t; \hat{t})$, at any fixed pseudo-time, $\hat{t}$. As such, beyond its use for physics inquiry exemplified above, SFR provides a basis for numerical solution of a coarse-grained representation of a flow (i.e., LES) with numerical grid resolution proportional to $\ell = \sqrt{2 \hat{\nu} \hat{t}}$. Equation \eqref{eq:aux-diffusion-residual-stress}, together with its formal solution in Eq.\ \eqref{eq:residual-stress-solution}, defines the residual stress tensor. The stress tensor requires modeling, because an exact solution of Eqs. \eqref{eq:aux-diffusion-residual-stress} or \eqref{eq:residual-stress-solution} requires knowledge of the flow field at equivalent resolution to DNS to accurately specify the velocity gradient down to scale $\hat{t} = 0$.

Any proposed model form may be substituted into Eq.\ \eqref{eq:aux-diffusion-residual-stress} to form an equation to determine model coefficients. For example, an eddy viscosity model may be assumed, with K41 scaling for the eddy viscosity,
\begin{equation}
    \sigma_{ij} = - 2 \nu_T S_{ij}
    \quad
    \text{where}
    \quad
    \nu_T = c_T \ell^{4/3}.
    \label{eq:eddy-viscosity-K41-1}
\end{equation}
In this case, with substitution, neglecting the spatial variation of the coefficient $c_T$ (i.e., $\nabla^2 c_T \approx 0$), the following equation is obtained,
\begin{equation*}
    \frac{2}{3} c_T (2\hat{\nu} \hat{t})^{-1/3} S_{ij}
    =
    - 2 (A_{ik} A_{jk})^{(d)}.
\end{equation*}
The proposed model form has a single scalar coefficient, $c_T$, yet is required to satisfy a tensor equation. Thus, a least-squares fit can minimize the error in the proposed model, resulting in
\begin{equation}
    \nu_T 
     =  \frac{3}{4} \frac{\Pi_1}{\| \mathbf{S} \|^2},
    \quad
    \sigma_{ij} = - 2 \nu_T S_{ij}.
    \label{eq:eddy-viscosity-K41}
\end{equation}
where $\Pi_1$ is given by Eq.\ \eqref{eq:cascade-rate-single-scale}.
The steps used to generate this model resemble those of the dynamic procedure of Germano \cite{Germano1991} and Lilly \cite{Lilly1992} that has become quite popular. However, the current SFR-based dynamic procedure uses Eq. \eqref{eq:aux-diffusion-residual-stress} in place of the Germano identity \cite{Germano1992}. The advantage of the current approach is that the model coefficient, $3/4$, can be determined analytically using pen-and-paper and does not require the calculation of a test filter. In this case, with the K41-scaled eddy viscosity assumption, the resulting model resembles the minimum dissipation model \cite{verstappen2011does}, except that the current model includes both strain-rate self-amplification and vortex stretching. A straightforward extension to anisotropic grids can be accomplished with a tensorial pseudo-viscosity having eigenvectors and eigenvalues corresponding to the grid orientation and aspect ratio, respectively. The resulting SFR-based dynamic model would resemble the anisotropic minimum dissipation model \cite{Rozema2015}, but with an additional inclusion of vorticity stretching.

\section{A New Class of SFR-Based Stress Models Using Residual Kinetic Energy}
\label{sec:SFR models}

Johnson \cite{Johnson2022Alternative} proposed and validated the performance of Eq.\ \eqref{eq:eddy-viscosity-K41} in HIT using (global) spatial averaging over homogeneous directions, as sometimes done with Germano-based dynamic models. The results confirmed that the analytical dynamic coefficient, $3/4$, provided an accurate dissipation rate with the resolved kinetic energy spectrum closely matching that of Gaussian-filtered DNS, see Eq.\ \eqref{eq:aux-diffusion-velocity-solution}. In this section, the SFR theory for LES modeling is extended to incorporate kinetic energy considerations, thereby eliminating the need for global spatial averaging. This is achieved by introducing theory-based stabilization for a range of SFR-based dynamic models, including a new one-equation model with an SFR-based transport equation for residual kinetic energy. These advancements, relative to Ref. \cite{Johnson2022Alternative}, will enable a careful and detailed study of the artificial bottleneck effect in LES.

\subsection{Eddy Viscosity Model via Local Equilibrium of Residual Kinetic Energy}
\label{subsec: dvisc}

The scaling of the eddy viscosity in Eq.\ \eqref{eq:eddy-viscosity-K41-1} results from dimensional analysis on the residual stress tensor assuming dependence on the resolution length scale consistent with Kolmogorov's 1941 theory with local equilibrium and scale-independent dissipation rate of (resolved) kinetic energy. Another possibility is to consider the eddy viscosity as a function of the residual kinetic energy, $e$, whose evolution in pseudo-time and physical time is given by Eqs.\ \eqref{eq:aux-diffusion-residual-energy} and \eqref{eq:residual-kinetic-energy-transport}, respectively. Assuming $\nu_T = \nu_T(\ell, e)$, dimensional analysis leads to,
\begin{equation}
    \nu_T = c_{\nu} \sqrt{e} \ell.
    \label{eq:eddy-viscosity-energy-1}
\end{equation}
Substituting this form into Eq.\ \eqref{eq:aux-diffusion-residual-stress}, 
and applying least-squares minimization, the following model is obtained,

\begin{equation}
    c_{\nu}
    =  \frac{  \ell \sqrt{e} }{\left( e + \frac{1}{2} \ell^2 \| \mathbf{A} \|^2 \right)}  \left(\ell \sqrt{e} \nabla^2 c_{\nu}  -  \frac{A_{ik} A_{jk}S_{ij}}{\| \mathbf{S} \|^2}  \right),
\quad
    \sigma_{ij} = -2 \nu_T S_{ij}.
    \label{eq:eddy-viscosity-energy}
\end{equation}


In Equation \eqref{eq:eddy-viscosity-energy}, the spatial variation of the coefficient $c_{\nu}$ is retained (i.e., $\nabla^2 c_\nu$).
The local value of the residual kinetic energy, $e$, may be determined by solving its transport equation, Eq.\ \eqref{eq:residual-kinetic-energy-transport}, with appropriate models for unclosed terms. Alternatively, a local equilibrium of residual kinetic energy may be assumed.

A local equilibrium of residual kinetic energy implies a local balance of sources and sinks in Eq.\ \eqref{eq:residual-kinetic-energy-transport}. If direct forcing at unresolved scales may be neglected, ${W}^{(5)} = 0$, local equilibrium implies
\begin{equation}
    \Pi = \epsilon.
    \label{eq:local-equilibrium}
\end{equation}
Where $\Pi = -\sigma_{ij} S_{ij}$ is the modeled production rate of residual kinetic energy and $\epsilon = {W}^{(4)}$ is the viscous dissipation rate of residual kinetic energy. In this approximation, {spatio-temporal} variation of $e$ is neglected. Likewise neglecting spatial variation of $e$ in the pseudo-time equation for residual kinetic energy, Eq.\ \eqref{eq:aux-diffusion-residual-energy}, results in
\begin{equation}
    \frac{\partial e_\eq}{\partial \hat{t}} = \hat{\nu} A_{ij} A_{ij},
    \hspace{0.1\linewidth}
    e_\eq(\mathbf{x}, t; \hat{t}=0) = 0.
    \label{eq:aux-diffusion-equilibrium-energy}
\end{equation}
Now, the local equilibrium assumption itself leads to scaling arguments via dimensional analysis and Kolmogorov's similarity hypotheses, thus,
\begin{equation}
    e_\eq = c_e \ell^{2/3}.
    \label{eq:equilibrium-energy-K41-1}
\end{equation}
Substitution of Eq.\ \eqref{eq:equilibrium-energy-K41-1} into Eq.\ \eqref{eq:aux-diffusion-equilibrium-energy}, with $\ell^2 = 2 \hat{\nu} \hat{t}$ leads to,
\begin{equation}
    e_\eq = \frac{3}{2} \ell^2 \| \mathbf{A} \|^2.
    \label{eq:equilibrium-energy-K41-2}
\end{equation}
{Then, substitution of this equilibrium residual kinetic energy}, Eq.\ \eqref{eq:equilibrium-energy-K41-2}, into Eq.\ \eqref{eq:eddy-viscosity-energy} yields,
\begin{equation}
     \nu_T
    =   \frac{3}{4} \left( \ell^2  \nabla^2 \nu_T+  \frac{\Pi_1}{\| \mathbf{S} \|^2}     \right).
    \label{eq:eddy-viscosity-energy-eqm}
\end{equation}
This result agrees with the previous K41 approximation, Eq. \eqref{eq:eddy-viscosity-K41}. The only difference is the retention of an additional term proportional to $\nabla^2 \nu_T$, which creates an elliptic differential equation for the eddy viscosity coefficient that acts as a local averaging operator on the viscosity. Such localized averaging (on the scale of $\ell$) acts as a stabilizing feature of the model that directly emerges from the SFR theory. (Note that, since we are here dealing with sufficiently inaccurate models, the substitution of a more computationally efficient local averaging method is appropriate and unlikely to significantly affect the model's performance.)

From the perspective of energy equilibrium, another consideration emerges. At locations where $\nu_T < 0$, the viscosity is negative and the modeled production rate of residual kinetic energy  $\Pi $ is negative. While this should occur at a strong minority of points in the domain, it should be appreciated that the above model is constructed with a pointwise assumption of equilibrium via $\Pi = \epsilon$, Eq.\ \eqref{eq:local-equilibrium}. Where $\nu_T < 0$, this is only possible if $\Pi = \epsilon = 0$, because $\epsilon$ is a non-negative quantity by definition. At such points in the modeled flow, the local equilibrium energy is thus $e_\eq = 0$.
That is, the local equilibrium assumption precludes the possibility of local backscatter, $\Pi < 0$, because that would lead to a unbalanced sink on the right-hand side of the residual kinetic energy transport equation, Eq.\ \eqref{eq:residual-kinetic-energy-transport}.  Therefore, following a consistent assumption of local equilibrium,
\begin{equation}
    e_\eq = \frac{3}{2} \ell^2 \| \mathbf{A} \|^2 ~\mathcal{H}(\nu_T).
    \label{eq:equilibrium-energy-K41}
\end{equation}
Where $\mathcal{H}$ denotes the Heaviside step function,
\begin{equation}
    \mathcal{H}(x) =
    \left\lbrace
    \begin{array}{l l}
        1 & \text{if}~ x > 0\\
        0 & \text{otherwise}
    \end{array}
    \right..
    \label{eq:Heaviside-step-function}
\end{equation}
Thus, an eddy viscosity model with consistent application of a local equilibrium (K41) assumption is,
\begin{equation}
     \sigma_{ij} = - 2  \mathcal{H}(\nu_T) \nu_T S_{ij},
    \quad 
    \Pi=2 \nu_T \| \mathbf{S} \|^2.
    \label{eq:eddy-viscosity-energy-eqm-2}
\end{equation}
In effect, this implements a clipping of the dynamic viscosity, $\nu_T$, which is a common trick used to stabilize a Germano-based dynamic eddy viscosity model. However, we are arguing here that the clipping in Eqs.\ \eqref{eq:equilibrium-energy-K41} and \eqref{eq:eddy-viscosity-energy-eqm-2} arises naturally as a consequence of local equilibrium, $e_\eq = 0$. The stabilization of the above SFR-based equilibrium eddy viscosity model is a consequence of the theory itself, not an \textit{ad hoc} imposition.

\subsection{Eddy Viscosity Model via Transport Equation for Residual Kinetic Energy}
\label{subsec:residual energy transport}

If it is desirable to capture departures of residual kinetic energy from local equilibrium and potentially capture local backscatter events, an additional transport equation can be considered to determine a non-equilibrium value of $e$ to be used with Eq.\ \eqref{eq:eddy-viscosity-energy}. To do this, a transport equation for residual kinetic energy is solved in the LES, from Eq.\ \eqref{eq:residual-kinetic-energy-transport},
\begin{equation}
    \frac{\partial e}{\partial t}
    + \frac{\partial}{\partial x_j}\left(
        U_j e
        - \nu \frac{\partial e}{\partial x_j}
        + q_j
    \right)
    =
    \Pi - \epsilon.
    \label{eq:modeled-residual-kinetic-energy-transport}
\end{equation}

Where direct forcing of small-scale energy, ${W}^{(5)}$ is assumed negligible. The production rate, $\Pi = - \sigma_{ij} S_{ij}$, does not require additional closure beyond what is needed for the resolved momentum equation, Eq. \eqref{eq:gen-Navier-Stokes}. Thus, two new modeling approximations are needed  \cite{Schumann1975}. First, the residual flux, $\mathbf{q}$, must be modeled. In most existing kinetic energy transport models, a gradient diffusion hypothesis is applied. This would certainly be commensurate with the use of an eddy viscosity model for the residual stress tensor, $\boldsymbol{\sigma}$. 
Second, a model for the viscous dissipation rate is required. The dissipation rate can be modeled as,
\begin{equation}
    \epsilon = C_\epsilon \frac{e^{3/2}}{\ell},
    \label{eq:dissipation-equation-formula}
\end{equation}
Then  enforcing $\Pi = \epsilon$ for $e = e_{eq}$ at equilibrium leads to,
\begin{equation}
    \epsilon = \left( \frac{e}{e_{eq}}\right)^{3/2} \Pi_{eq}.
\end{equation}
Where $\Pi_{eq}$ represents the modeled production rate of residual kinetic energy at equilibrium, Eq.\ \eqref{eq:eddy-viscosity-energy-eqm-2}. The modeled dissipation rate may be thought as a nonlinear relaxation toward equilibrium that follows standard dissipation scaling. Furthermore, the residual flux $\mathbf{q}$ is evaluated via a gradient diffusion hypothesis, 
\begin{equation}
    q_j   = - \alpha_T \frac{\partial e}{\partial x_j}
    \label{eq:residual-flux-model},
\end{equation}
Where $\alpha_T$ is the eddy diffusivity, modeled as,
\begin{equation}
    \alpha_T = c_{\alpha} \sqrt{e} \ell.
    \label{eq:eddy-diffusivity-energy-1}
\end{equation}
Then adding Eq.\ \eqref{eq:residual flux-1} and  Eq.\ \eqref{eq:residual flux-2} leads to,
\begin{equation}
    \frac{\partial q_j}{\partial \ell^{2}} =
    \frac{1}{2} \nabla^2 q_j
    +   \frac{\partial U_j}{\partial x_k} \left(\frac{\partial e}{\partial x_k}+ \frac{\partial P}{\partial x_k} \right)
    +\frac{\partial U_i}{\partial x_k} \frac{\partial \sigma_{ij}}{\partial x_k} ,
    \label{eq:residual-total-flux-evolution}
\end{equation}
Substitution of Eq.\ \eqref{eq:residual-flux-model} and Eq.\ \eqref{eq:eddy-diffusivity-energy-1} into Eq.\ \eqref{eq:residual-total-flux-evolution} with further simplifications and applying least-squares minimization, the following $\alpha_T$ model is obtained,

\begin{equation}
    c_{\alpha}= \frac{  \ell \sqrt{e} }{\left( e + \frac{1}{2} \ell^2 \| \mathbf{A} \|^2 \right)}  
    \left(\ell \sqrt{e} \nabla^2 c_{\alpha}  - 2 \frac{\frac{\partial e}{\partial x_j} S_{ij} \frac{\partial e}{\partial x_k} }
    {\| \nabla{e} \|^2} \right)
    \label{eq:gradient-diffusion-hypothesis}
\end{equation}
where $-\frac{\partial e}{\partial x_j} S_{ij} \frac{\partial e}{\partial x_k}$ depends on how the strain rate is aligned with the unresolved energy gradient vector, which is conceptually similar to the strain-rate self-amplification and vortex stretching, Eq. \ref{eq:resolved energy cascade}.
The negative sign accounts for the fact that scalar gradients preferentially align with compressive strain-rate eigenvectors \cite{Ashurst1987}, so that the diffusivity given by Eq.\ \eqref{eq:gradient-diffusion-hypothesis} will be predominantly positive throughout the simulated flow.

\subsection{Mixed Model via Local Equilibrium of Residual Kinetic Energy}

Mixed models are built by combining a structure-based nonlinear model and a functional eddy viscosity model to provide better alignment with the actual stress tensor as proved via filtered DNS (a priori testing),
\begin{equation}
\sigma_{ij}=\sigma^{(1)}_{ij}+ \sigma^{(2)}_{ij}= \ell^2 (A_{ik} A_{jk})^{(d)}  -2 \nu_T S_{ij},
    \label{eq:mixed-model}
\end{equation}
The first component, $\boldsymbol{\sigma}^{(1)}$, comes from the resolved part of the formal solution to Eq.\ \eqref{eq:aux-diffusion-residual-stress}, see Eq.\ \eqref{eq:cascade-rate-single-scale}. The eddy viscosity component, $\boldsymbol{\sigma}^{(2)}$, models the unresolved portion; see Eq. \eqref{eq:cascade-rate-multi-scale}.
From dimensional analysis, $\nu_T = \nu_T(\ell, e)$ following Eq.\ \eqref{eq:eddy-viscosity-energy-1}.
Substituting this form into Eq.\ \eqref{eq:aux-diffusion-residual-stress}, 
and applying least-squares minimization, the eddy viscosity is obtained,
\begin{equation}
   \nu_T   = 
    \frac{e\ell^2}{e+\frac{1}{2}\ell^4 \| \mathbf{B} \|^2 }\left( \nabla^2 \nu_T   -\ell^2 \frac{B_{mkl}B_{nkl}S_{mn}}{\| \mathbf{S} \|^2}  \right).
    \label{eq:dmixed eddy-viscosity-energy-eqm}
\end{equation}
Where $B_{ijk} =\partial^2 U_i/ \partial x_j \partial x_k$ is the Hessian of the resolved velocity. (The appearance of second derivatives is mathematically similar to the test filter calculation mentioned in Ref. \cite{chapelier2018coherent, hasslberger2023dynamic} where a small scale activity sensor is constructed by evaluating the ratio of the test-filtered to grid-filtered velocity gradient tensor. If the two filters are close together, the leading-order term in Taylor series expansion would contains second derivatives of the velocity field. On a related note, in the previous two eddy viscosity models in section \ref{subsec: dvisc} and \ref{subsec:residual energy transport}, a scale-dependent eddy viscosity coefficient \cite{Porte-Agel2000, BouZeid2005} could be sought instead of the scale similarity assumptions; this would also lead to second-derivatives.) The residual kinetic energy can be expressed as the energy contribution of each model separately,
\begin{equation}
e = e^{(1)}+e^{(2)}= \frac{1}{2} \ell^2 \| \mathbf{A} \|^2 + e^{(2)}.
\end{equation}
Where $e^{(1)}= \frac{1}{2} \ell^2 \| \mathbf{A} \|^2$ is the {residual kinetic energy} due to the resolved part of the formal solution to Eq.\ \eqref{eq:aux-diffusion-residual-energy}, i.e., the energy related to the nonlinear gradient stress component, and $e^{(2)}$ is associated with the eddy viscosity component. The scale-wise derivative of $e^{(1)}$ is,
\begin{equation}
    \frac{\partial e^{(1)}}{\partial \ell^{2}} = \frac{1}{2} \| \mathbf{A} \|^2 +\frac{\ell^{2}}{2} \left( \frac{1}{2}  \nabla^2   \| \mathbf{A} \|^2  - B_{ijk}B_{ijk} \right).
    \label{eq: residual kinetic energy(1)}
\end{equation}
To find $e^{(2)}$, we refer back to Eq. \eqref{eq:aux-diffusion-residual-energy}, and using Eq. \eqref{eq: residual kinetic energy(1)} leads to,
\begin{equation}
    \frac{\partial e^{(2)}}{\partial \ell^{2}} = \frac{1}{2}  \nabla^2 e^{(2)} + \frac{1}{2} \ell^2  B_{ijk}B_{ijk}.
    \label{eq: residual kinetic energy(2)}
\end{equation}
The equilibrium argument for $e^{(2)}$  follows the same K41 scaling  as Eq. \eqref{eq:equilibrium-energy-K41-1}, yields to,
\begin{equation}
    e^{(2)}_\eq = \frac{3}{2} \ell^4 \| \mathbf{B} \|^2,
    \label{eq:equilibrium-energy-K41-3}
\end{equation}
Thus, substitution of Eq. \eqref{eq:equilibrium-energy-K41-3}, into Eq. \eqref{eq:dmixed eddy-viscosity-energy-eqm} leads to,

\begin{equation}
    \nu_T  = 
    \frac{3\ell^2}{4}\left( \nabla^2 \nu_T   -\ell^2 \frac{B_{mkl}B_{nkl}S_{mn}}{\| \mathbf{S} \|^2}  \right),
    \quad
     \sigma_{ij}
     = \ell^2 (A_{ik} A_{jk})^{(d)} - 2 \mathcal{H}(\nu_T) \nu_T S_{ij}
    \label{eq:mixed-model-energy-eqm}
\end{equation}
The modeled production rate of residual kinetic energy is obtained from Eq. \eqref{eq:mixed model production rate},
\begin{equation}
    \Pi = - \sigma_{ij} S_{ij}
    = \ell^2 A_{ik} A_{jk} S_{ij}+2 \nu_T \| \mathbf{S} \|^2 =\Pi_1 +\Pi_2
    \label{eq:mixed model production rate}
\end{equation}
Local equilibrium for $e^{(2)}$ implies,
\begin{equation}
    \Pi_2 = \epsilon_2,
    \label{eq:local-equilibrium-mixed}
\end{equation}
At locations where $\nu_T < 0$,  $\Pi_2 $ is negative, and according to the local equilibrium argument Eq. \eqref{eq:local-equilibrium-mixed}, this requires clipping of $e^{(2)}$ via $\Pi_2 = \epsilon_2 = 0$, inline with Section \ref{subsec: dvisc}. The local equilibrium value at such points is thus $e^{(2)}_\eq = 0$.






\section{Results}\label{sec:results}

In this section, the artificial bottleneck effect that leads to a spectral overshoot is inspected in \textit{a posteriori} LES results. The numerical methods utilized are briefly described before a detailed comparison of various residual stress models. Then, the effect of filter width, grid resolution, and Reynolds number arepresented.

\subsection{Numerical Methods}
\label{sec:numerics}

Direct Numerical Simulations (DNS) of homogeneous isotropic turbulence (HIT) are carried out to provide baseline results for comparison with SFR-based models. The incompressible Navier-Stokes equation, Eq.\ \eqref{eq:Navier-Stokes}, is solved using a pseudo-spectral method in a triply-periodic domain of size $(2\pi)^3$. 
The pressure maintains a divergence-free velocity field as a projection tensor applied to the other right-hand side terms in the Navier-Stokes equation. 
Time advancement is accomplished with a second-order Adams-Bashforth scheme. 
The non-linear terms are dealiased using the $2\sqrt{2}/3$ rule with phase-shifting.

The body force is activated for only the first two wave-shells are forced at each time step with magnitude linearly proportional to the velocity. The coefficient of proportionality is determined at each time step to maintain constant kinetic energy spectrum for the first two wave-shells.
In this case the initial velocity field is chosen using a random number generator to create one instance from a Gaussian random field ensemble satisfying the divergence-free condition.
The initial velocity field is set to have a model energy spectrum, 
and {the simulation is allowed to run for $6$ large-eddy turnover times to allow for the solution to become statistically representative of fully developed turbulence before statistics are recorded for $6$ more turnover times}.
For this paper, DNS at $Re_\lambda \approx 400$ (scale separation of $L / \eta \approx 460$, where $L$ is the integral length scale) is used with grid resolution $\kappa_{\max} \eta = 1.4$ using $1024^3$ collocation points. Important parameters for the DNS simulation are given in Table \ref{tab:DNS Simulation parameters }.

The Large-Eddy Simulations (LES) are computed for HIT with the same numerical approach as above with a few modifications. Of course, the residual stress models are activated for LES according to the equations laid out in Section \ref{sec:SFR models}. Moreover, the non-linear terms both in the momentum and subgrid equations are dealiased using the $2/3$ rule (instead of phase-shift dealiasing). The LES averages are computed over $\sim 20$ large-eddy turnover times. The DNS results are filtered with a Gaussian kernel for direct comparison with the LES results from each model.

\begin{table}[h]
    \centering
    \setlength{\tabcolsep}{10pt}
    \begin{tabular}{c c c c c c c c}
        \hline 
        N & $Re_\lambda$ &  $\epsilon$ &  $\nu$ &
        $\eta$ &  $\tau_\eta$ & $\Delta t$ & $k_{max}\eta$ \\
        \hline
         $1024^3$ & $395$ & $0.113$ & $2\times 10^{-4}$  & $2.89\times 10^{-3}$  & $0.042$& $2 \times 10^{-4}$ &$1.4$  \\ 
        \hline
    \end{tabular}
    \caption{Numerical details for the DNS simulation used in this paper}
    \label{tab:DNS Simulation parameters }
\end{table}

\subsection{The Artificial Bottleneck Effect at $Re_\lambda = 400$ and $\ell/\eta = 24$}
\begin{figure}[htb]
\begin{minipage}{0.5\textwidth}
  \begin{tikzpicture}
  \node (img)  {\includegraphics[scale=1]{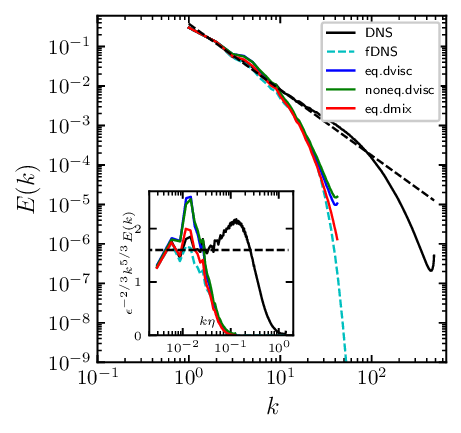}};
    \node[above=of img, node distance=0cm, xshift=-3.5cm, yshift=-1.5cm,font=\color{black}] {(a)};
  \end{tikzpicture}
\end{minipage}%
\begin{minipage}{0.5\textwidth}
  \begin{tikzpicture}
  \node (img)  {\includegraphics[scale=1]{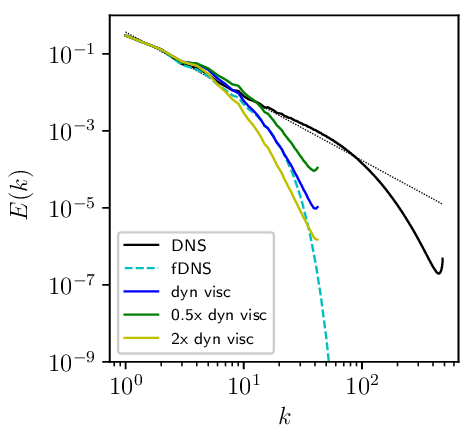}};
  \node[above=of img, node distance=0cm, xshift=-3.5cm, yshift=-1.5cm,font=\color{black}] {(b)};
  \end{tikzpicture}
\end{minipage}%
\caption{ Energy spectra for a posteriori testing: (a) using three different models  with $\ell=24 \eta$ at $\kappa_{max} \ell=3.0$, and (b) using the equilibrium eddy viscosity model with different model coefficient. A Kolmogorov
spectrum, $E(\kappa)=1.6 \; \varepsilon^{\frac{2}{3}} \kappa^{-\frac{5}{3}} $ is shown for reference in both panels. The inset in panel (a)
shows the premultiplied spectrum on a log–linear plot.}
\label{fig:sfig1}
\end{figure}

First, an \textit{a posteriori} comparison of the artificial bottleneck effect in SFR-based dynamic LES models in forced isotropic turbulence at $Re_{\lambda} \approx 400$  is carried out with a filter width $\ell / \eta=24$, which is at the small-scale end of the inertial range identified in the \textit{a priori} tests \cite{Johnson2022Alternative}.
The SFR-based models tested are: (i) equilibrium eddy viscosity (eq.dvisc), Eq.\ \ref{eq:eddy-viscosity-energy-eqm-2}, (ii) equilibrium mixed model (eq.dmix), Eq.\ \ref{eq:mixed-model-energy-eqm}, and (iii) non-equilibrium eddy viscosity (noneq.dvisc), Eqs.\ \ref{eq:eddy-viscosity-energy} \& \ref{eq:modeled-residual-kinetic-energy-transport}.
Using a numerical resolution of $\kappa_{max} \ell=3.0$ for the LES models, Figure \ref{fig:sfig1}(a) compares the energy spectra from DNS, Gaussian-filtered DNS, and the SFR-based models introduced in this paper. 
The models shown here are all SFR-based dynamic models, meaning that the model coefficients are determined from an identical dynamic procedure using Stokes flow regularization (SFR) instead of the well known Germano dynamic procedure \cite{Germano1991}. 
It is emphasized that the SFR-based dynamic procedure provides a pen-and-paper derivation of the dynamic coefficient and does not require a test filter. Additionally, the coefficients are calculated locally without requiring averaging along homogeneous directions. The difference in computational cost for these models is briefly explored in Appendix A. 

Figure \ref{fig:sfig1}(a) demonstrates that the SFR-based dynamic models produce LES spectra that roll off at the correct wavenumber, indicating the success of the dynamic procedure in determining an accurate model coefficient for each model.
To further emphasize the success of the SFR-based dynamic procedure in this regard, Figure \ref{fig:sfig1}(b) shows that changing the SFR-based dynamic eddy viscosity coefficient $3/4$ in Eq. \eqref{eq:eddy-viscosity-energy-eqm} to $(0.5\times 3/4= 3/8)$ and $(2\times 3/4= 3/2)$, which shifts the roll-off to the right and left. That is, increasing the coefficient above the SFR-based value introduces too much dissipation into the LES model and causes the spectrum to roll off at a lower wavenumber. On the other hand, decreasing the coefficient below its SFR-based value reduces the dissipation and causes the roll-off of the spectrum to occur at a higher wavenumber. This demonstrates that the SFR-based dynamic procedure produces a coefficient that provides an accurate amount of dissipation to the LES.

The inset of Figure \ref{fig:sfig1}(a) shows the pre-multiplied energy spectra of the SFR models. The observed spectral overshoot produced by the SFR-based eddy viscosity models is with a wavenumber and amplitude similar to that produced by the dynamic Smagorinsky model based on Germano-Lilly dynamic procedure \cite{Germano1991, Lilly1992} in the inset of Figure \ref{fig:bottleneck}(b). 
It is thus evident that the artificial bottleneck effect is not caused by a deficiency in the dynamic procedure for the model coefficient, nor by under- or over-prediction of the energy dissipation rate.
The introduction of history/memory effects through the residual kinetic energy tranport equation does not impact the artificial bottleneck effect, as observed in the results of the non-equilibrium dynamic eddy viscosity model.
However, the height of the spectral bump is reduced in the dynamic mixed model results compared to the eddy viscosity models. The observation suggests that the introduction of a nonlinear gradient component in the residual stress closure improves the interaction between the residual stress model and the resolved energy cascade in LES.

The nonlinear gradient component, $\ell^2 A_{ik} A_{jk}$, of the dynamic mixed model, Eq.\ \eqref{eq:mixed-model}, represents the direct calculation of the resolved single-scale component of the energy cascade in Eq.\ \eqref{eq:resolved energy cascade}, $\Pi_1 = \Pi_{s1} + \Pi_{\omega 1}$.
To understand the relative success of the dynamic mixed model in reducing the magnitude of the spectral overshoot, the SFR-based LES models are compared with filtered DNS in terms of the following two quantities: 
\begin{equation}
    s^\ast=-\frac{\sqrt{6} S_{ij} S_{jk} S_{ki} }{ (S_{mn}S_{mn})^{3/2} } = \frac{\Pi_{s1}}{|\Pi_{s1,\text{max}}|},  \hspace{0.05\linewidth}
     \omega^\ast=\frac{\sqrt{6} \Omega_i S_{ij} \Omega_j }{ 2 \Omega_k \Omega_k (S_{mn}S_{mn})^{1/2} } = \frac{\Pi_{\omega1}}{|\Pi_{\omega1,\text{max}}|}.
     \label{eq:flow topology}
\end{equation}
The $s^\ast$ parameter quantifies the efficiency of the single-scale strain-rate self-amplification contribution to the energy cascade, $ -1 \leq s^\ast \leq 1 $, based on the ratio of the three strain-rate eigenvalues. The most efficient flow structure, $s^* = 1$, is given by two equal positive (stretching) eigenvalues and one negative (compressing) eigenvalue with twice the magnitude.  Note that $s^*$ is equivalent to the Lund-Rogers parameter \cite{Lund1994} applied to the resolved strain-rate tensor.
The $\omega^\ast$ parameter indicates the efficiency of the single-scale vorticity stretching contribution to the cascade, $ -1 \leq \omega^\ast \leq 1$, based on the alignment of the vorticity vector with the three strain-rate eigenvectors (and their associated eigenvalues). The most efficient flow structure for the vorticity stretching, $\omega^* = 1$, is given by parallel alignment of the vorticity vector with the strain-rate eigenvector corresponding to one positive eigenvalue, with the other two strain-rate eigenvalues being equal with half the magnitude of the first. Further discussion of cascade efficiencies can be found in Refs.\ \cite{Ballouz2018Tensor, Johnson2021Role}.

\begin{figure}[htb]
\begin{minipage}{0.5\textwidth}
  \begin{tikzpicture}
  \node (img)  {\includegraphics[scale=1]{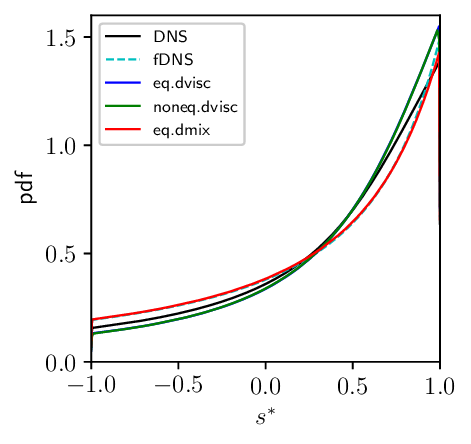}};
    \node[above=of img, node distance=0cm, xshift=-3.5cm, yshift=-1.5cm,font=\color{black}] {(a)};
  \end{tikzpicture}
\end{minipage}%
\begin{minipage}{0.5\textwidth}
  \begin{tikzpicture}
  \node (img)  {\includegraphics[scale=1]{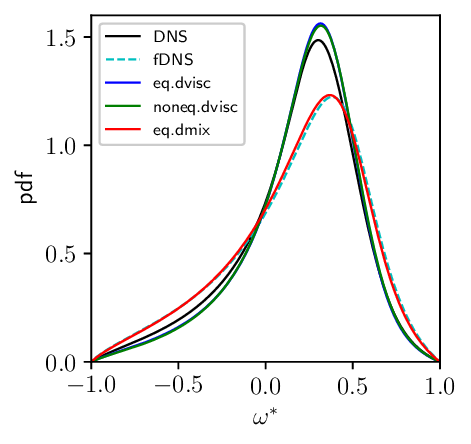}};
  \node[above=of img, node distance=0cm, xshift=-3.5cm, yshift=-1.5cm,font=\color{black}] {(b)};
  \end{tikzpicture}
\end{minipage}%
\caption{PDF of  $\omega^ \star $ and $s^ \star$, Eq. \eqref{eq:flow topology}, for LES with three different models compared with DNS and filtered DNS.}
\label{fig:sfig2}
\end{figure}

Figure \ref{fig:sfig2} compares the PDFs of $s^\ast$ and $\omega^\ast$ produced by the SFR-based LES models compared with DNS and filtered DNS results.
The DNS and filtered DNS provide qualitatively similar distributions of $s^*$ and $\omega^*$ because the same restricted Euler dynamics (coming from nonlinear self-advection) are active in both cases. 
Turbulence tends to favor positive values (forward cascade) for both, but the vortex stretching process tends to favor less efficient flow topologies (based on alignment of vorticity with the strain-rate eigenvectors).
However, there are noticeable quantitative differences in the PDFs, particular for $\omega^*$.
As one might expect, the eddy viscosity models (both equilibrium and non-equilibrium) produce flow topology statistics similar to those seen in DNS. The interaction of an eddy viscosity in LES with the resolved energy cascade is evidently similar to the interaction of the cascade with viscous dissipation in DNS, even if the LES models do allow for spatial variation of the viscosity.
In contrast, the mixed model is in better agreement with filtered DNS distributions of $s^*$ and $\omega^*$, presumably due to its ability to more accurately represent the tensorial structure of the residual stress.
Thus, the reduction in the spectral overshoot by the dynamic mixed model in Fig. \ref{fig:sfig1}(a) can be explained by its ability to more faithfully represent the structural details of the energy cascade, especially near the resolution scale $\ell$.


Table \ref{tab:HIT-stats} provides a quantitative comparison of the models in terms of common statistics of interest for homogeneous isotropic turbulence (HIT). In addition to the SFR-based models, a class of traditional LES models are tested and shown alongside the newly proposed models for comparison: (i) The constant Smagorinsky model (Smag), (ii) The dynamic Smagorinsky model with clipping (dSmag.clip), and (iii) the dynamic Smagorinsky model with global averaging  (dSmag.avg). The coefficient for the static Smagorinsky model (Smag) is tuned to match the wavenumber at which the filtered DNS spectra rolls off, to simulate a ``perfect'' dynamic Smagorinsky model (with global coefficient) while the coefficient for dSmag.clip and dSmag.avg are obtained through a dynamic procedure \cite{Germano1991, Lilly1992}. Also included are the SFR-based models with globally averaged coefficients \cite{Johnson2022Alternative}. 

The first column of the table compares the normalized resolved dissipation rate, $\langle S_{ij} S_{ij} \rangle \tau_\eta^2$, which is over-predicted by all the eddy viscosity models compared to the filtered DNS results due to the artificial bottleneck effect. 
Given that the resolved dissipation rate can be calculated by integrating the spectrum weighted by $\kappa^2$, the nature of these over-predictions can be inferred from the energy spectra.
As expected, the mixed model provides a better prediction.
The next two columns compare the LES prediction of resolved strain-rate self-amplification ($\Pi_{s1}$) and resolved vortex stretching ($\Pi_{\omega 1}$), as defined in Eq.\ \eqref{eq:resolved energy cascade}, as a fraction of the total energy cascade rate. 
All the eddy viscosity models show a significant over-prediction of these quantities, with a larger percent error compared with the aforementioned $\langle S_{ij} S_{ij} \rangle$ over-prediction, attributable to the fact that the energy cascade rates are a triple product of velocity gradients.
Together, these first three columns demonstrate that the eddy viscosity models tend to over-predict the magnitude of velocity gradients in the resolved field despite their accurate prediction of the roll-off wavenumber, Fig.\ \ref{fig:sfig1}. This observation underscores some of the consequences of the artificial bottleneck effect in addition to the $\sim 10\%$ over-prediction of resolved kinetic energy for this case.

The final three columns compare LES predictions for skewness and flatness values of longitudinal and transverse filtered velocity gradient components. Recall that the skewness and flatness of a Gaussian random variable are $0$ and $3$, respectively, and the skewness of tranverse velocity gradients is zero by symmetry arguments, i.e., $\langle A_{12}^3 \rangle = 0$. The filtering at $\ell = 24 \eta$ significantly reduces the non-Gaussianity of the DNS results (here, at $Re_\lambda \approx 400$).
Most LES models generally produce realistic skewness and flatness values for the filtered velocity gradient, except for the eddy viscosity model with globally averaged coefficient (dvisc.avg) \cite{Johnson2022Alternative}. This suggests that the spatial variation of the eddy viscosity is important for reproducing the correct amount of intermittency in the resolved scales. The development of SFR-based dynamic model stabilization in Section \ref{sec:SFR models} to enable local determination of the eddy viscosity shows important advantages in this respect.




\begin{table}[h]
    \centering
    \begin{tabular}{c c c c c c c}
        \hline
        Model & 
        $\left\langle S_{ij} S_{ij} \right\rangle \tau_\eta^2$ &
        $\left\langle \Pi_{s1} \right\rangle / \left\langle \Pi \right\rangle$ &
        $\left\langle \Pi_{\omega 1} \right\rangle / \left\langle \Pi \right\rangle$ &
        $\left\langle A_{11}^3 \right\rangle / \left\langle A_{11}^2 \right\rangle^{3/2}$ &
        $\left\langle A_{11}^4 \right\rangle / \left\langle A_{11}^2 \right\rangle^{2}$ &
        $\left\langle A_{12}^4 \right\rangle / \left\langle A_{12}^2 \right\rangle^{2}$ \\
        \hline
         unfiltered DNS (DNS) & 0.5 & -- & -- &$-0.59$  & $8.1$  & $12.7$ \\ 
        filtered DNS (fDNS) & $1.7$e-$2$ & $0.37$ & $0.12$ & $-0.42$ & $3.9$ & $4.7$  \\
        
        SFR eq. mixed (eq.dmix)  & $1.8$e-$2$  & $0.40$ & $0.13$ & $-0.39$ & $3.8$ & $4.6$\\
        
        SFR mixed (dmix.avg)  \cite{Johnson2022Alternative} &  $1.9$e-$2$& $0.45$ & $0.15$ & $-0.39$ & $3.9$ & $4.8$  \\
        SFR eq. eddy viscosity (eq.dvisc) &  $2.2$e-$2$& $0.53$ & $0.18$ & $-0.41$ & $3.6$ & $4.5$\\
        SFR eddy viscosity (dvisc.avg) \cite{Johnson2022Alternative} &  $1.9$e-$2$& $0.50$ & $0.17$ & $-0.46$ & $4.8$ & $6.8$  \\
        SFR non-eq.eddy viscosity (noneq.dvisc) &  $2.3$e-$2$ & $0.58$ & $0.19$ & $-0.40$ & $3.7$ & $4.8$ \\
        Smag &  $2.3$e-$2$ & $0.58$ & $0.19$ & $-0.44$ & $3.6$ & $5.0$ \\
        dynamic Smagorinsky (dSmag.clip) &  $2.2$e-$2$ & $0.59$ & $0.20$ & $-0.44$ & $3.8$ & $4.7$ \\
        dynamic Smagorinsky (dSmag.avg) &  $2.2$e-$2$ & $0.58$ & $0.19$ & $-0.43$ & $3.7$ & $5.1$\\
        \hline
    \end{tabular}
    \caption{ Statistical comparison of LES models with Gaussian-filtered DNS at  $\ell=24 \eta$ with $128^3$ resolution ($\kappa_{max} \ell=3.0$). The abbreviation eq. refers to the local equilibrium assumption, Eq.\ \eqref{eq:local-equilibrium}.  }
    \label{tab:HIT-stats}
\end{table}


\subsection{Effect of Filter Width}
\label{sec:Artificial bottle neck effect}
\begin{figure}[htb]
\begin{subfigure}{1\textwidth}
\begin{minipage}{0.5\textwidth}
  \begin{tikzpicture}
  \node (img)  {\includegraphics[scale=1]{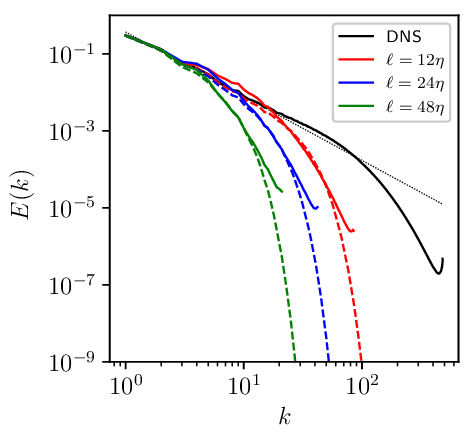}};
    \node[above=of img, node distance=0cm, xshift=-3.5cm, yshift=-1.5cm,font=\color{black}] {(a)};
  \end{tikzpicture}
\end{minipage}%
\begin{minipage}{0.5\textwidth}
  \begin{tikzpicture}
  \node (img)  {\includegraphics[scale=1]{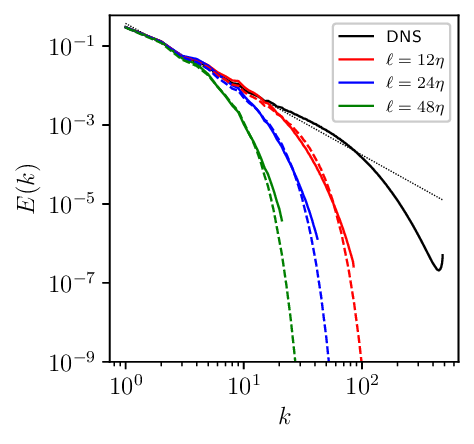}};
    \node[above=of img, node distance=0cm, xshift=-3.5cm, yshift=-1.5cm,font=\color{black}] {(b)};
  \end{tikzpicture}
\end{minipage}%
\end{subfigure}
\caption{Energy spectra for a posteriori testing at $\kappa_{max} \ell=3.0$: (a) using equilibrium eddy viscosity, 
and (b) using equilibrium mixed model. The dashed lines represent the Gaussian-filtered DNS. A Kolmogorov spectrum, $E(\kappa)=1.6 \; \varepsilon^{\frac{2}{3}} \kappa^{-\frac{5}{3}} $ is shown for reference in both panels.}
\label{fig:sfig3}
\end{figure}

The \textit{a posteriori} LES results in the previous section were shown at a filter width of $\ell = 24 \eta$ with a grid resolution of $\kappa_{\max} \ell = 3.0$ for isotropic turbulence at $Re_\lambda = 400$. To further test the SFR-based models, results at different filter scales are discussed in this section. Figure \ref{fig:sfig3} shows the energy spectra comparison at $\kappa_{max} \ell=3.0$ for the equilibrium eddy viscosity and mixed models at different filter scales, $\ell$, compared {with the DNS energy spectrum filtered at $\ell$}. 
The non-equilibrium eddy viscosity model was also compared in this way (not shown), and its results were very similar to the equilibrium eddy viscosity model shown here.
As expected, the SFR-based models successfully capture the change in roll-off wavenumber as the filter scale is varied. 
The spectral overshoot is observed for the eddy viscosity results in Figure \ref{fig:sfig3}(a) for $\ell = 12\eta$ and less-so for $\ell = 48\eta$. The mixed model reduces the magnitude of the spectral overshoot for all filter widths. This is explored quantitatively in Table \ref{tab: resolved dissipation comparison_400_Re}.

First, the resolved dissipation rate, $\langle S_{ij} S_{ij} \rangle$, is over-predicted by all models tested at all filter widths shown in Table \ref{tab: resolved dissipation comparison_400_Re}. The middle column for $\ell = 24 \eta$ is the same data from Table \ref{tab:HIT-stats}, renormalized in terms of the inertial range timescale $\tau_{\ell} = \epsilon^{-1/3} \ell^{2/3}$.
The model predictions generally improve with decreasing filter width, except for the constant Smagorinsky model (Smag) where the coefficient was tuned to match the spectrum roll off. The improvement in the models, particularly for $\ell = 12\eta$, appears to stem from potential cancellation effects in the prediction of the energy spectrum, Fig.\ \ref{fig:sfig3}. That is, the LES spectra at $\ell = 12 \eta$ show alternative over-prediction and under-prediction compared with the filtered DNS results. 

Upon closer inspection of Figure \ref{fig:sfig3}, it is possible to detect a slight change in the model performance as $\ell$ becomes closer to $\eta$ (and further from $L$) and enters the viscous range of scales. Namely, the model results show a spectrum that falls off at a wavenumber that is slightly {below} the filtered DNS for $\ell = 12\eta$. This indicates that the self-similar (inertial range) scaling assumptions used to derive the dynamic model coefficients are less accurate if $\ell$ is not sufficiently large compared to $\eta$. One way to address this shortcoming could be to introduce scale-dependence into the dynamic coefficient calculation, which for the Germano-based filtering approach requires a second test filter \cite{Porte-Agel2000, BouZeid2005}. In the SFR formulation, this would require the use of higher-order derivatives of the velocity field {(e.g., second derivatives of the velocity for the equilibrium eddy viscosity model)}.

\begin{table}[h]
    \begin{center}
    \begin{tabular}{c c c c }
        & $\ell=0.005L_{box}$ & $\ell=0.01L_{box}$&$\ell=0.02L_{box}$ \\
        \hline
        Filter width & $\ell=12\eta$ & $\ell=24\eta$ & $\ell=48\eta$ \\
        \hline
         Model &   $\left\langle S_{ij} S_{ij} \right\rangle \tau_\ell^2$  &   $\left\langle S_{ij} S_{ij} \right\rangle \tau_\ell^2$  & $\left\langle S_{ij} S_{ij} \right\rangle \tau_\ell^2$   \\
        \hline
        fDNS &  $1.22$ & $1.14$ & $1.13$  \\
        
        eq.dmix &  $1.21$ & $1.23$ & $1.25$   \\
        eq.dvisc &  $1.41$ & $1.49$ & $1.53$  \\
        noneq.dvisc &  $1.51$  & $1.59$ & $1.63$ \\
        Smag &  $1.54$ & $1.53$ & $1.48$ \\
        dSmag.clip &  $1.46$ & $1.54$ & $1.58$ \\
        dSmag.avg &  $1.46$ & $1.51$ & $1.55$ \\
        \hline
    \end{tabular}
    \caption{Statistical comparison of SFR-based models resolved dissipation rate with Gaussian-filtered DNS and Smagorinsky models  at $\kappa_{max} \ell= 3.0$ with different filter width $\ell$ at $Re_{\lambda}=400$.{ $L_{box}$ is the simulation domain box size ($2\pi$) where $L/L_{box}= 0.21$ ($L$ is the integral length scale),  and $\tau_\ell / \tau_\eta=(5.24, 8.33, 13.2) $ for DNS respectively}. The energy spectrum comparison for \textit{a posteriori} testing is shown in Figure \ref{fig:sfig3}. }
    \label{tab: resolved dissipation comparison_400_Re}
    \end{center}
\end{table}

Perhaps more so than the magnitude of velocity gradient fluctuations, the prediction of kinetic energy is an important criteria in \textit{a posteriori} testing of LES models. Of course, both the kinetic energy and the resolved dissipation rate can be calculated as integrals of the energy spectrum with different weighting, so an error in one metric may in principle show up as an error in the other.
Thus, just as the artificial spectral bump leads to an over-prediction of the resolved dissipation rate, so also the LES models over-predict the resolved kinetic energy, $\langle E\rangle$, as seen in Table \ref{tab: kin energy comparison_400_Re} .
For applications, of course, it is more important for LES to accurately predict the total kinetic energy, $\langle T \rangle = \langle E \rangle + \langle e \rangle$, than any filtered quantity.
Therefore, Table \ref{tab: kin energy comparison_400_Re} shows each of these three quantities (total, resolved, and residual kinetic energy) for the LES results compared with filtered DNS.
Of course, the total kinetic energy, $\langle T \rangle = 0.73$, is established by unfiltered DNS and does not depend on the filter.
The table shows that the over-prediction in resolved kinetic energy leads to a similar over-prediction of residual kinetic energy (due to the model's assumption of equilibrium and inertial range scaling) and thus an over-prediction of the total kinetic energy in the flow.
This is particularly true for the eddy viscosity models, although it should be noted that Smagorinsky models make no prediction of the residual kinetic energy unless supplemented with, e.g., a Yoshizawa model. One benefit of SFR theory is that it naturally provides a prediction of the residual kinetic energy, even for algebraic (equilibrium) models, e.g., Eq.\ \eqref{eq:equilibrium-energy-K41} for the dynamic eddy viscosity model.

The over-prediction of kinetic energy in the eddy viscosity models is relatively similar at $\ell = 12\eta$ and $\ell = 24 \eta$. At $\ell = 48 \eta$, there is less over-prediction in the eddy viscosity models, which can be attributed to a smaller magnitude of the spectral overshoot, see Fig. \ref{fig:sfig3}(a). At this coarsest filter width, only about $70\%$ of the total kinetic energy is resolved in the filtered velocity field, so the energy cascade is less robust. This may explain why the bottleneck effect is less evident.
These over-predictions of kinetic energy may be traced, thus, to the artificial bottleneck effect seen in Fig.\ \ref{fig:sfig1}.
As expected, the mixed model shows a smaller over-prediction of kinetic energy that improves (slightly) as the filter scale decreases.

Note that the kinetic energy over-prediction in the eddy viscosity models grows as $\ell$ decreases while the resolved dissipation rate over-prediction decreases. 
This is because there is cancellation of errors at $\ell = 12 \eta$ where the bottleneck causes over-prediction of energy at lower wavenumbers but the spectrum rolls off too soon so there is under-prediction of energy at higher wavenumbers, but the energy at such high wavenumbes is too low, so the cancellation of errors will not be important, see Figure \ref{fig:sfig3}. However, the cancellation of errors plays a role in the resolved dissipation rate calculated via $\int_{0}^{\kappa_{\max}} \kappa^2 E(\kappa) d\kappa$ where the energy discrepancy at higher wavenumber is more influential compared to the energy calculation.

\begin{table}[H]
    \begin{center}
    \begin{tabular}{c|c c c|c c c|c c c}
        \multirow{2}{*}{Filter width} &
        \multicolumn{3}{c}{$\ell=0.005L_{box}$} & \multicolumn{3}{|c|}{$\ell=0.01L_{box}$} & \multicolumn{3}{c}{$\ell=0.02L_{box}$} \\
        \multirow{1}{*}{} & 
        \multicolumn{3}{c}{$\ell=12\eta$} & \multicolumn{3}{|c|}{$\ell=24\eta$} & \multicolumn{3}{c}{$\ell=48\eta$} \\
        \cline{1-10}
         Model & $\langle e \rangle$ &  $\langle E \rangle$  & $\langle T \rangle$ & $\langle e \rangle$ & $\langle E \rangle$  & $\langle T \rangle$ & $\langle e \rangle$ & $\langle E \rangle$  & $\langle T \rangle$ \\
        \hline
        fDNS &  $0.07$ & $0.66$ & $0.73$ & $0.12$ & $0.61$ & $0.73$ & $0.20$ & $0.53$ & $0.73$   \\
        
        eq.dmix &  $0.09$ & $0.69$ & $0.78$ & $0.15$ & $0.64$ & $0.79$ & $0.22$ & $0.56$ & $0.78$   \\
        eq.dvisc &  $0.11$ & $0.74$ & $0.85$ & $0.16$ & $0.67$ & $0.83$ & $0.24$ & $0.57$ & $0.81$ \\
        noneq.dvisc &  $0.11$ & $0.74$ & $0.85$  &  $0.17$ & $0.68$ & $0.85$ &  $0.25$ & $0.58$ & $0.83$    \\
        Smag &  $-$ & $0.73$ & $-$  &  
        $-$ & $0.67$ & $-$ & $-$ & $0.57$ & $-$     \\
        dSmag.clip&   $-$ & $0.73$ & $-$  & 
        $-$ & $0.68$ & $-$ & $-$ & $0.56$ & $-$     \\
        dSmag.avg&   $-$ & $0.73$ & $-$  &  
        $-$ & $0.67$ & $-$ & $-$ & $0.56$ & $-$     \\
        \hline
    \end{tabular}
    \caption{Statistical comparison of SFR-based models kinetic energy with Gaussian-filtered DNS and Smagorinsky models at $\kappa_{max} \ell= 3.0$ with different filter width $\ell$ at $Re_{\lambda}=400$; $L_{box}$ is the box size. The energy spectrum comparison of both eddy viscosity and mixed models in a posteriori testing is shown in Figure \ref{fig:sfig3}. }
    \label{tab: kin energy comparison_400_Re}
    \end{center}
\end{table}

In summary, the artificial bottleneck in LES leads to an over-prediction of the resolved turbulent kinetic energy by $\sim 10\%$ for eddy viscosity models. The resolved velocity gradient magnitudes are over-predicted by and even greater extent, leading to more over-prediction of the residual kinetic energy. In the end, an over-prediction of $\sim 15\%$ in total kinetic energy is observed for these cases when an eddy viscosity model is used. This over-prediction descrease to $\sim 3\%$ for resolved kinetic energy and $\sim 6\%$ for total kinetic energy when the dynamic mixed model is used. Given that the LES resolution scale is usually set proportionally to the integral length scale rather than the viscous scale, the artificial bottleneck represents a potentially important feature of LES model inaccuracies even at high Reynolds numbers.

\subsection{Effect of Numerical Resolution}

Next, the numerical resolution  relative to the filter width $\kappa_{max} \ell$ is varied for the LES models. Figure \ref{fig:kmax_ell=1.5,6.0} shows the energy spectra for the SFR-based equilibrium eddy viscosity model at different $\kappa_{max} \ell$ for under-resolved LES ($\kappa_{max} \ell= 1.5$)  and over-resolved LES ($\kappa_{max} \ell= 6.0$). 
When under-resolved, the models continue to produce reasonable results but do not capture the roll-off of the spectrum.
The spectral overshoot is observed at all resolutions, highlighting the fact that it is a feature of modeling errors in the residual stress, not numerical discretization errors.

The over-resolved LES simulation captures very low energy modes deep into the roll-off of the spectrum. 
There is some additional error in the shape of the spectral roll-off that becomes more apparent (compared with the Gaussian-filtered DNS) at these high wavenumbers, but it is relatively insignificant because the energy content at this region is too low. 
The non-equilibrium eddy viscosity and equilibrium mixed models show an energy pile-up for the over-resolved case ($\kappa_{max} \ell= 6.0$), which can be mitigated by increasing the local averaging coefficient (i.e., the coefficient in front of the Laplacian term) in Eq. \eqref{eq:eddy-viscosity-energy} and Eq. \eqref{eq:mixed-model-energy-eqm} (not shown).

\begin{figure}[htb]
    \includegraphics[scale=1]{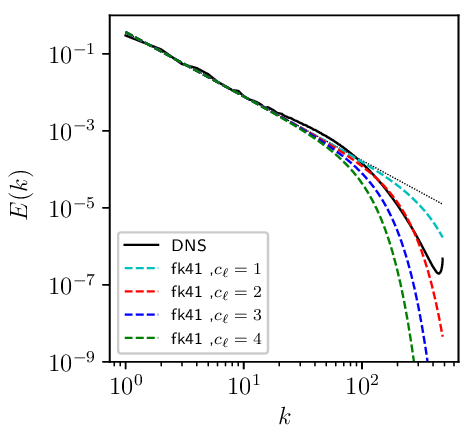}
    \caption{Filtered K41 spectrum at different filter width $\ell = c_\ell \; \eta$. A Kolmogorov spectrum, $E(\kappa)=1.6 \; \varepsilon^{\frac{2}{3}} \kappa^{-\frac{5}{3}} $ is shown for reference.}
    \label{fig:min acceptable resolution}
\end{figure}

When considering the effect of grid resolution at a fixed filter width, it is interesting to document a minimum recommended resolution. For DNS, a resolution of $\kappa_{\max} \eta$ between $1$ and $2$ is generally considered minimally sufficient for capturing the viscous energy dissipation (though perhaps not higher-order moments of the velocity gradient). 
The above results suggest a grid resolution of $\kappa_{\max} \ell \approx 3.0$.
To explore this further, an ideal infinite-Re spectrum can be filtered at a length scale $\ell = c_\ell \eta$ and compared with the (unfiltered) DNS spectrum.
The filtered ideal infinite-Re spectrum is given as,
\begin{equation}
\overline{E}(\kappa)=1.6 \; \varepsilon^{\frac{2}{3}} \kappa^{-\frac{5}{3}} \exp(-\kappa^2 \ell^2).
\label{eq:filtered k-41 spectrum}
\end{equation}
Figure \ref{fig:min acceptable resolution} shows different filtered K41 spectrum (fk41) at different choices of filter width $c_\ell$. In terms of matching the roll-off wavenumber of the spectrum, the best agreement with the DNS occurs at $c_\ell \approx 2$. Using this result, the LES resolution of $k_{max} \ell = 3.0$ is comparable to a DNS resolution of $\kappa_{max} \eta \approx  1.5$, which is a typical value.{ Note that a resolution of $\kappa_{\max} \ell = 3.0$, when using 2/3-rule dealiasing, translates to a grid spacing (i.e., distance between collocation points)} of {$\Delta x = 0.71 \ell$}. Note that our SFR-based definition of the filter scale ($\ell$) differs from the traditional filter width (e.g. Ref.\ \cite{Borue1998}) by a factor of $\Delta = \sqrt{20} \ell$, so our finding corresponds to $\Delta x = 0.16 \Delta$. The traditional filter width definition ($\Delta$) is based on the width of a tophat filter with second moment equal to that of the Gaussian filter, but SFR theory defines the filter width ($\ell$) directly based on the second moment of the Gaussian (i.e., variance of a Gaussian PDF). At coarser grid resolutions (relative to the filter width), numerical errors will play a significant role in determining model performance.

\begin{figure}[htb]
\begin{minipage}{0.5\textwidth}
  \begin{tikzpicture}
  \node (img)  {\includegraphics[scale=1]{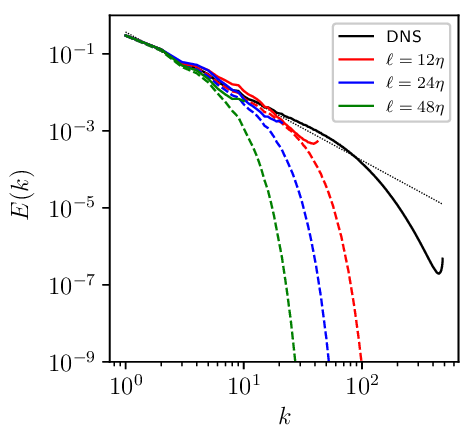}};
    \node[above=of img, node distance=0cm, xshift=-3.5cm, yshift=-1.5cm,font=\color{black}] {(a)};
  \end{tikzpicture}
\end{minipage}%
\begin{minipage}{0.5\textwidth}
  \begin{tikzpicture}
  \node (img)  {\includegraphics[scale=1]{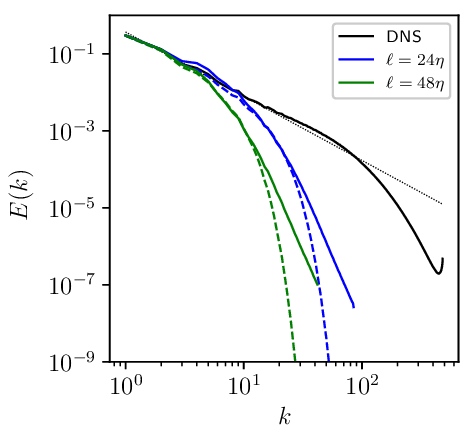}};
  \node[above=of img, node distance=0cm, xshift=-3.5cm, yshift=-1.5cm,font=\color{black}] {(b)};
  \end{tikzpicture}
\end{minipage}%
\caption{Energy spectra for a posteriori testing using equilibrium eddy viscosity model : (a) under-resolved LES with $\kappa_{max} \ell= 1.5$, and (b) over-resolved LES with $\kappa_{max} \ell= 6.0$. In (a,b) the dashed lines represent the Gaussian-filtered
DNS. A Kolmogorov
spectrum, $E(\kappa)=1.6 \; \varepsilon^{\frac{2}{3}} \kappa^{-\frac{5}{3}} $ is shown for reference in both panels.}
\label{fig:kmax_ell=1.5,6.0}
\end{figure}


\subsection{ Visualization of Vortical Structures}
\label{App:B}

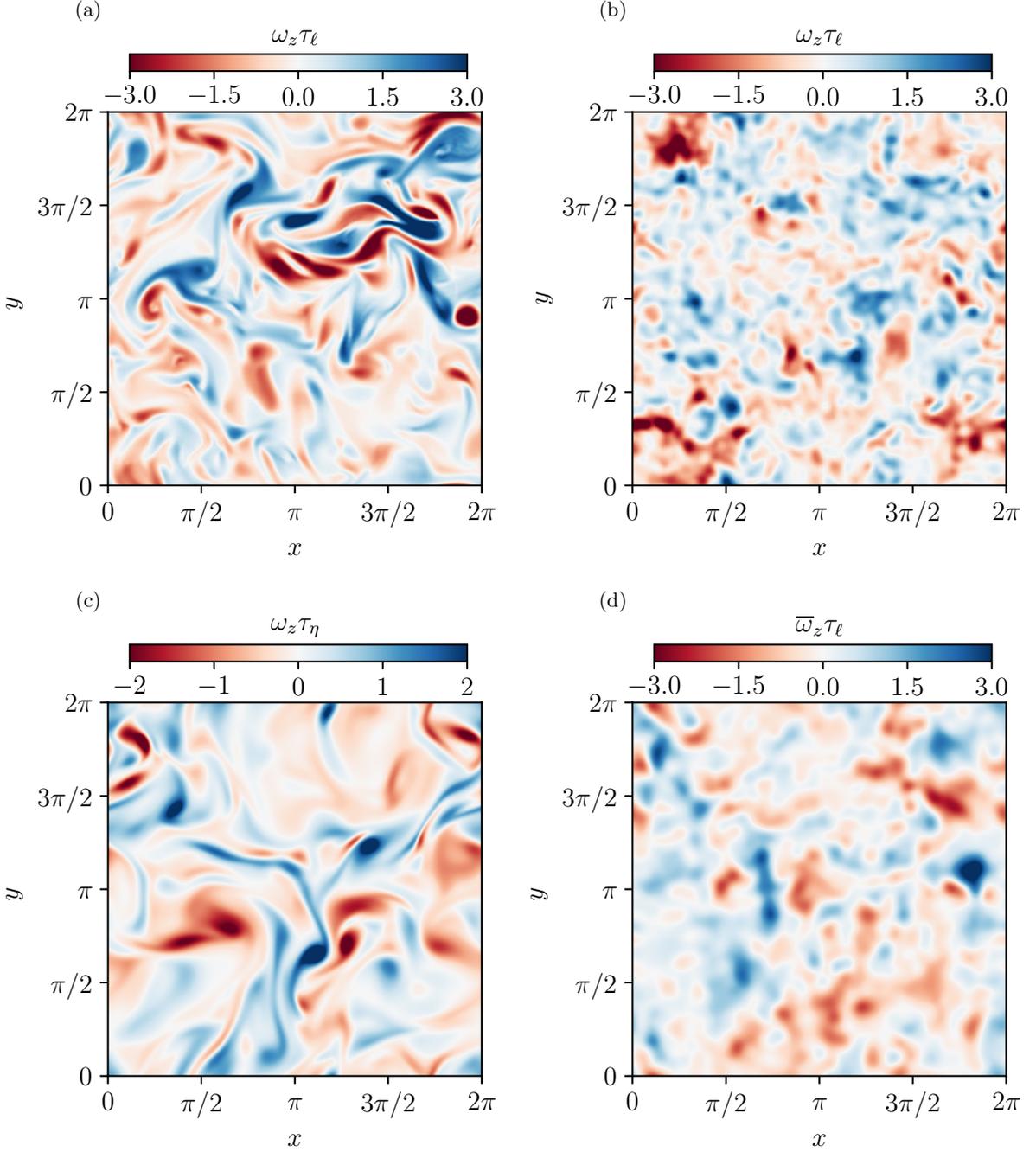
\begin{figure}[htb]
\begin{minipage}{0.45\textwidth}
\centering
  \begin{tikzpicture}
  \node (img)  {\includegraphics[width=\textwidth]{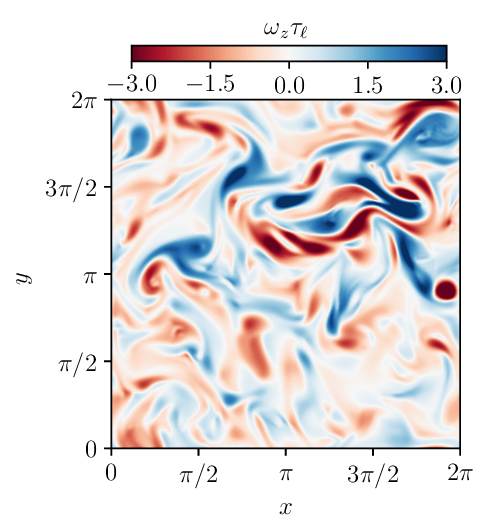}};
    \node[above=of img, node distance=0cm, xshift=-2.5cm, yshift=-1.5cm,font=\color{black}] {(a)};
  \end{tikzpicture}
\end{minipage}%
\begin{minipage}{0.45\textwidth}
\centering
  \begin{tikzpicture}
  \node (img)  {\includegraphics[width=\textwidth]{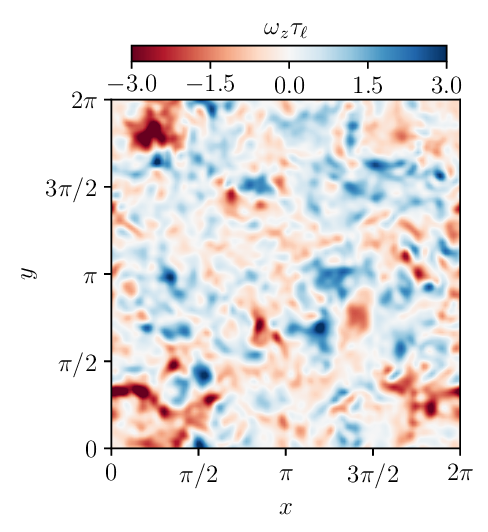}};
  \node[above=of img, node distance=0cm, xshift=-2.5cm, yshift=-1.5cm,font=\color{black}] {(b)};
  \end{tikzpicture}
\end{minipage}%
  \hfill
\begin{minipage}{0.45\textwidth}
\centering
  \begin{tikzpicture}
  \node (img)  {\includegraphics[width=\textwidth]{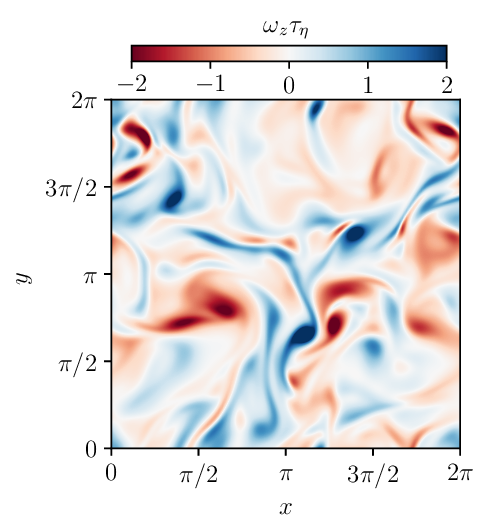}};
    \node[above=of img, node distance=0cm, xshift=-2.5cm, yshift=-1.5cm,font=\color{black}] {(c)};
  \end{tikzpicture}
\end{minipage}%
\begin{minipage}{0.45\textwidth}
\centering
  \begin{tikzpicture}
  \node (img)  {\includegraphics[width=\textwidth]{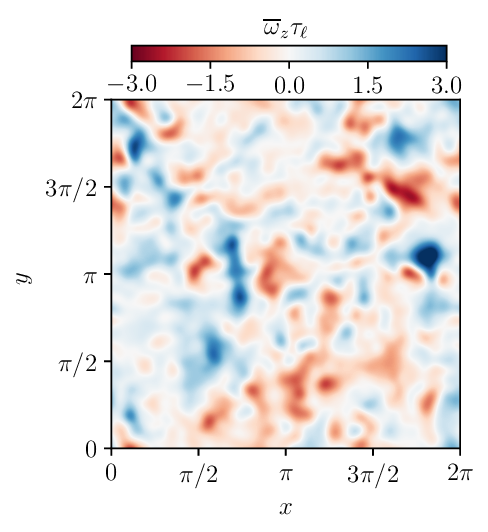}};
    \node[above=of img, node distance=0cm, xshift=-2.5cm, yshift=-1.5cm,font=\color{black}] {(d)};
  \end{tikzpicture}
\end{minipage}%
\caption{ The z component of vorticity along  xy plane  in the HIT simulation: (a) equilibrium eddy
viscosity model with  $\ell=24 \eta$ at  $\kappa_{max} \ell=6.0$, (b) equilibrium mixed model with  $\ell=24 \eta$ at  $\kappa_{max} \ell=3.0$, (c) unfiltered DNS at $Re_\lambda=65$, and (d) Gaussian-filtered DNS with  $\ell=24 \eta$ at $Re_\lambda=400$.}
\label{fig:slices}
\end{figure}

Figure \ref{fig:slices} illustrates the out-of-plane vorticity on a 2D slice, comparing the SFR-based LES models (the equilibrium eddy viscosity model and the mixed model) against the Gaussian-filtered DNS with a filter width $\ell=24 \eta$ at $Re_\lambda=400$ and DNS at a lower Reynolds number ($Re_\lambda=65$). The vorticity field has been renormalized using the inertial range timescale $\tau_{\ell} = \epsilon^{-1/3} \ell^{2/3}$. Figure \ref{fig:slices}(b) shows that the mixed model produces flow structures (e.g., vortex tubes) with strong qualitative similarity to those of the filtered DNS, Figure \ref{fig:slices}(d). In contrast, the eddy viscosity model produces a significantly different flow structure containing more shear layer-like structures (i.e., vortex sheets rather than vortex tubes). The flow structures from the eddy viscosity LES seen in Figure \ref{fig:slices}(a)  closely resemble those observed in a lower Reynolds number DNS in Figure \ref{fig:slices}(c) inline with the observation in Figure \ref{fig:sfig2}, both of which differ substantially from the filtered DNS.
The flow structures seen in simulations using the other eddy viscosity models such as the classical dynamic Smagorinsky and the non-equilibrium SFR-based are very similar to those in Figure \ref{fig:slices}(a).

The shear layers produced by the eddy viscosity models are potentially associated with the inverse cascade vortex thinning mechanism in 2D turbulence, suggesting a possible connection between the (artificial) bottleneck effect in 3D turbulence and the inverse cascade in 2D turbulence \cite{Johnson2021Role}. In combination with Figure \ref{fig:sfig2}, these visualizations provide further evidence that the mixed model is able to diminish the artificial bottleneck effect because of its more accurate representation of flow structure and energy cascade dynamics near the filter scale.
While the \textit{a priori} advantages of mixed models are well-known, these results suggest that mixed models can also provide significant \textit{a posteriori} advantages which should be considered, particularly as one moves to more complex flow regimes. 

\subsection{Infinite Reynolds number simulations}

At $Re_\lambda = 400$, the trends in LES predictions with respect to changing the filter width may be caused by two potential effects. First, as $\ell / L$ decreases, and a larger percent of the kinetic energy is resolved. On the other hand, as $\ell / \eta$ decreases, viscous effects become resolved and the self-similar (inertial range) scaling assumptions built into the dynamic procedure for the SFR-based models become inaccurate. To distinguish between these two effects, it is useful to run the LES models at infinite Reynolds number.
While it precludes a direct comparison with (filtered) DNS, it can be useful to test the SFR-based LES models at infinite $Re$ ($\nu = 0$) to check the models performance at $\ell/L \ll 1$ together with $\ell / \eta = \infty$, in an inertial range where the flow is approximately independent of large scale forcing and there are no viscous effects.

\begin{figure}[htb]
\begin{minipage}{0.5\textwidth}
  \begin{tikzpicture}
  \node (img)  {\includegraphics[scale=1]{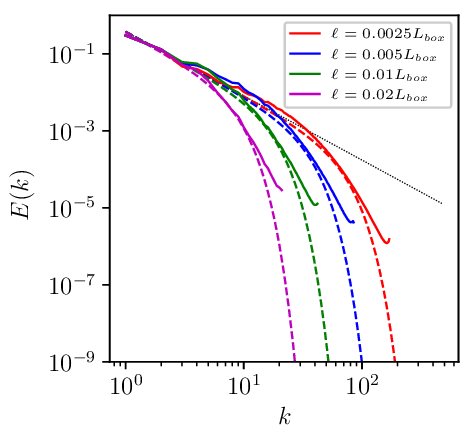}};
    \node[above=of img, node distance=0cm, xshift=-3.5cm, yshift=-1.5cm,font=\color{black}] {(a)};
  \end{tikzpicture}
\end{minipage}%
\begin{minipage}{0.5\textwidth}
  \begin{tikzpicture}
  \node (img)  {\includegraphics[scale=1]{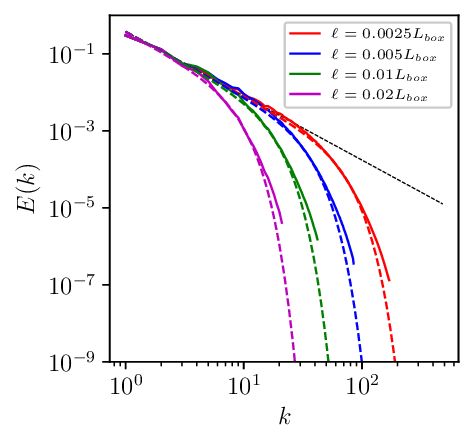}};
  \node[above=of img, node distance=0cm, xshift=-3.5cm, yshift=-1.5cm,font=\color{black}] {(b)};
  \end{tikzpicture}
\end{minipage}%
\caption{Infinite Reynolds number energy spectra for a posteriori testing at $\kappa_{max} \ell= 3.0$ : (a) using equilibrium eddy viscosity model, and (b) using equilibrium mixed model. A Kolmogorov spectrum, $E(\kappa)=1.6 \; \varepsilon^{\frac{2}{3}} \kappa^{-\frac{5}{3}} $ is shown for the ideal infinite-Re spectrum in both panels.}
\label{fig:inf_Re_kmax_ell=3}
\end{figure}

Figure \ref{fig:inf_Re_kmax_ell=3} shows the comparison between the ideal infinite-Re filtered spectrum and the LES models at different filter width $\ell$.
The grid resolution in each LES case is set to $\kappa_{\max} \ell = 3.0$.
The spectral bump stemming from an artificial bottleneck effect is similar to the finite-Re case and is especially noticeable in the eddy viscosity models.
The mixed model continues to show a significantly reduced bottle neck effect.
This confirms that the molecular viscosity ($\nu$) is not involved in causing the observed spectral bump.
The spectral bump shifts to larger wavenumbers when the filter scale is reduced, suggesting that it happens at particular wavenumbers normalized by the filter width, $\kappa \ell$.
Comparing Figure \ref{fig:inf_Re_kmax_ell=3} with Figure \ref{fig:sfig3}, it may be observed that the roll-off wavenumber is slightly over-predicted by the LES models when run at infinite Reynolds number. This shows that the results in Figure \ref{fig:sfig3} are slightly influenced by the viscosity, most notably for $\ell = 12 \eta$.
The SFR-based dynamic coefficients appear to be slightly under-dissipative based on the infinite Re results.
The authors have observed that the dynamic Smagorinsky model also shows a similar under-dissipative behavior (data not shown).

Furthermore, a statistical comparison of the normalized resolved dissipation rate $\left\langle S_{ij} S_{ij} \right\rangle \tau_\ell^2$ at infinite $Re$ is shown in Table  \ref{tab: resolved dissipation comparison_inf_Re }. Table \ref{tab: resolved dissipation comparison_inf_Re } demonstrates that the resolved dissipation rate over-prediction decreases as $\ell$ decreases in line with the finite $Re$ results in Figure \ref{fig:sfig3} and Table \ref{tab: resolved dissipation comparison_400_Re}. The model predictions generally improve with decreasing filter width, though this trend is not as strong as in the finite Reynolds number case in Table \ref{tab: resolved dissipation comparison_400_Re} which had the benefit of error cancellation effects at different wavenumbers.
In the comparison between the tuned Smagorinsky model and the eddy viscosity models, the Smagorinsky model gives more accurate predictions across all scales. This advantage stems from the fact that the coefficient of the Smagorinsky model has been meticulously tuned, through trial and error, to match  the wavenumber at which the ideal filtered spectrum rolls off. Conversely, the dynamic eddy viscosity models display a slight tendency towards under-dissipation across all filter scales. This disparity underscores the effectiveness of the finely-tuned coefficient in the Smagorinsky model, and a slight under-prediction of the coefficient by the SFR-based dynamic procedure.

\begin{table}[H]
    \begin{center}
    \begin{tabular}{c c c c  c }
        \hline
         Filter width& $\ell=2.5$e-$3 L_{box}$ & $\ell=0.005L_{box}$ & $\ell=0.01L_{box}$ & $\ell=0.02L_{box}$  \\
        \hline
         Model &   $\left\langle S_{ij} S_{ij} \right\rangle \tau_\ell^2$  &   $\left\langle S_{ij} S_{ij} \right\rangle \tau_\ell^2$  & $\left\langle S_{ij} S_{ij} \right\rangle \tau_\ell^2$&  $\left\langle S_{ij} S_{ij} \right\rangle \tau_\ell^2$  \\
        \hline
        fDNS &  $1.08$ & $1.08$ & $1.07$  &  $1.05$   \\
        
        eq.dmix &  $1.21$ & $1.23$ & $1.25$     &  $1.26$    \\
        eq.dvisc &  $1.51$ & $1.53$ & $1.55$   &  $1.55$  \\
        noneq.dvisc &  $1.49$ & $1.53$ & $1.57$ &   $1.64$   \\
        Smag &  $1.34$ & $1.37$ & $1.39$     &  $1.48$  \\
        dSmag.clip &  $1.56$ & $1.58$ & $1.58$ &  $1.57$  \\
        dSmag.avg &  $1.53$ & $1.55$ & $1.55$     &  $1.55$  \\
        \hline
    \end{tabular}
    \caption{Statistical comparison of SFR-based models resolved dissipation rate $\left\langle S_{ij} S_{ij} \right\rangle \tau_\ell^2$  with the ideal infinite-Re filtered spectrum and Smagorinsky models at $\kappa_{max} \ell= 3.0$ with different filter width $\ell$ at $Re_{\lambda}=\infty $; $L_{box}$ is the box size. }
    \label{tab: resolved dissipation comparison_inf_Re }
    \end{center}
\end{table}

\section{Conclusions}\label{sec:conclusions}

This paper investigates the artificial bottleneck effect in large-eddy simulations (LES) of turbulent flows. In Navier-Stokes turbulence, the bottleneck effect is a well-known physical phenomenon related to the interaction of the kinetic energy cascade and viscous dissipation, leading to a pile up of energy at scales slightly larger than the Kolmogorov scale. The same effect is observed to occur in LES when an eddy viscosity model is used, leading to $\sim 10\%$ error in resolved kinetic energy and $15\%$ error in total kinetic energy for the cases analyzed.

Stokes Flow Regularization (SFR), a physics-inspired coarsening technique originally introduced to analyze the kinetic energy cascade \cite{Johnson2020Energy, Johnson2021Role, Johnson2022Alternative}, is herein extended to incorporate detailed considerations related to the residual kinetic energy, leading to the development of stabilization techniques that emerge directly from the SFR theory and allow for the local determination of dynamic coefficients.
This also leads to the introduction of a one-equation model based on residual kinetic energy transport using SFR theory, which captures some backscatter and other non-equilibrium effects.
The present extention of SFR theory for LES modeling facilitates a fair comparison between different model types with respect to the artificial bottleneck effect and over-prediction of kinetic energy.

\textit{A posteriori} tests of eddy viscosity and mixed models in homogeneous isotropic turbulence at $Re_\lambda = 400$ indicate that the mixed model significantly reduces the artificial bottleneck effect by providing an improved representation of grid-scale topology and resolved energy cascade efficiencies compared to the eddy viscosity models. The study also examined the effects of filter width, numerical resolution, and Reynolds number. Tests at infinite Reynolds number suggest that the bottleneck effect arises from modeling errors in the residual stress tensor rather than viscous effects or numerical discretization errors.

Even in the simple case of isotropic turbulence, the interaction between the resolved energy cascade and residual stress modeling errors is not straightforward. This type of effect is expected to be more important for more complex flows, where errors in closure approximations play an even more central role in the accuracy of the LES prediction. This motivates further research expanding SFR theory to include boundary effects and wall models, particle-laden and multi-phase flows, and density-stratified flows, for example.

\section*{Acknowledgements}
This material is based upon work supported by the National Science Foundation under Grant No.\ CBET-2152373.

\bibliography{references}

 \appendix
 \section{A Comparison of Computational Cost for SFR-based and Traditional LES Models}
 \label{App:A}
We conducted a computational cost analysis comparing SFR models to traditional LES models like the dynamic Smagorinsky model and reported the time required to run $1000 \Delta t$ in Table \ref{tab:Computational_cost}. The time is normalized by the the time needed to calculate LES using the constant Smagorinsky model, which serves as the fastest model for reference.
As expected, the test filter calculation in the dynamic Smagorinsky model is more computationally expensive than our SFR eddy viscosity models. However, the mixed model proves to be the most expensive due to the additional tensor contractions of second derivatives. It is important to note that this comparison is specific to the pseudo-spectral method used, and the relative computational cost may be quite different for finite-difference, finite-volume, and finite-element codes. Pseudo-spectral methods typically incur lower computational costs for the enforcement of the zero-divergence condition on the velocity field while calculating more accurate spatial derivatives compared to physical-domain solvers. However, the calculation of nonlinear products in a pseudo-spectral method requires a fast Fourier transform and dealiasing, which will generally require additional computational cost compared with physical-domain methods.

 \begin{table}[h]
    \centering
    \begin{tabular}{c c}
        \hline
        Model  & $1000 \Delta t$ \\
        \hline
        SFR eq. mixed (eq.dmix) &  $3.8$\\
        SFR eq. eddy viscosity (eq.dvisc) & $1.2$ \\
       SFR non-eq.eddy viscosity (noneq.dvisc) & $2.2$\\
        Smag & $1.0 $\\
        dynamic Smagorinsky (dSmag.clip) & $3.1$\\
        \hline
    \end{tabular}
   \caption{ Computational cost  comparison of LES models normalized by the static Smagorinsky  at  $\ell=24 \eta$ with $128^3$ resolution ($\kappa_{max} \ell=3.0$).}
    \label{tab:Computational_cost}
\end{table}

\end{document}